\pdfoutput=1

\documentclass[sigconf, nonacm]{acmart}
\RequirePackage{textcase}
\makeatletter
\expandafter\def\expandafter\@secfont\expandafter{\@secfont\MakeTextUppercase}
\makeatother

\AtBeginDocument{%
  }

\usepackage{amsmath}
\usepackage{tikz}
\usetikzlibrary{shapes.geometric,arrows.meta,positioning,fit,backgrounds,calc}
\usepackage{algorithm}
\usepackage{algpseudocode}
\algrenewcommand{\algorithmicfunction}{\textbf{Function}}
\makeatletter
\apptocmd{\fs@ruled}{%
  \def\@fs@pre{{\color{black}\hrule height.8pt depth0pt}\kern2pt}%
  \def\@fs@post{\kern2pt{\color{black}\hrule}\relax}%
  \def\@fs@mid{\kern2pt{\color{black}\hrule}\kern2pt}%
}{}{}
\makeatother
\usepackage{bbm}
\usepackage{enumitem}
\usepackage{booktabs}
\usepackage{tabularx}
\usepackage{multirow}
\usepackage{graphicx}
\graphicspath{{figures/}}
\usepackage{listings}
\usepackage[table]{xcolor}
\definecolor{semOpKw}{RGB}{20,90,170}     %
\definecolor{semStr}{RGB}{0,110,60}       %
\definecolor{semComment}{RGB}{120,120,120}%
\definecolor{semCol}{RGB}{170,60,30}      %
\definecolor{semFrame}{RGB}{200,200,200}
\lstdefinelanguage{semop}{
  morekeywords={sem_filter,sem_join,sem_topk,sem_map,sem_agg},
  keywordstyle=\bfseries\color{semOpKw},
  morecomment=[l]{\#},
  commentstyle=\itshape\color{semComment},
  morestring=[b]",
  stringstyle=\color{semStr},
  moredelim=[s][\color{semCol}\bfseries]{\{}{\}},
  sensitive=true,
}
\usepackage{tikz}
\usetikzlibrary{arrows.meta,positioning,shapes.geometric,calc,fit,backgrounds,shadows.blur}
\usepackage[most]{tcolorbox}
\tcbset{
  finding/.style={
    enhanced, breakable,
    colback=gray!7, colframe=gray!45,
    boxrule=0.5pt, arc=3pt,
    left=6pt, right=6pt, top=4pt, bottom=4pt,
  }
}

\tikzset{
  pipe/.style={
    draw=black, line width=0.6pt, rounded corners=2pt,
    font=\footnotesize, align=center,
    inner sep=4pt, minimum height=7mm,
  },
  stage/.style={
    pipe, fill=black!4, minimum width=28mm,
  },
  stageaccent/.style={
    pipe, fill=blue!6, draw=blue!55!black, minimum width=28mm,
  },
  io/.style={
    pipe, fill=white, minimum width=18mm,
  },
  data/.style={
    draw=black, line width=0.5pt, dashed, rounded corners=2pt,
    font=\scriptsize\itshape, align=center, inner sep=3pt,
  },
  ar/.style={
    -{Stealth[length=4pt,width=3.5pt]}, line width=0.55pt,
  },
  arf/.style={
    -{Stealth[length=4pt,width=3.5pt]}, line width=0.55pt, dashed,
  },
  lbl/.style={font=\scriptsize, align=center, inner sep=1pt},
}

\newcommand{\secref}[1]{\S\ref{#1}}
\newcommand{\condr}[1]{#1}
\definecolor{RED}{rgb}{1,0,0}%
\newif\iftrackchanges \trackchangesfalse
\ifdefined\cleanbuild \trackchangesfalse \fi
\DeclareRobustCommand{\deltxt}[1]{}
\DeclareRobustCommand{\cuttxt}[1]{}

\ifdefined\cleanbuild
  \newcommand{\delb}[1]{}
  \newcommand{\addr}[1]{#1}
\else
  \newcommand{\delb}[1]{{\color{blue}#1}}
  \newcommand{\addr}[1]{#1}
\fi
\makeatletter
\patchcmd{\@printendtopmatter}{\par\bigskip}{\par\smallskip}{}{}
\makeatother
\begin{document}

\title[PROBE: Learned \underline{Pro}xy-\underline{B}ased \underline{E}xecution for Highly Selective Multimodal Semantic Queries]{PROBE: Learned \underline{Pro}xy-\underline{B}ased \underline{E}xecution for Highly Selective Multimodal Semantic Queries}

\author{Haikai Zhao}
\affiliation{%
  \institution{Simon Fraser University}
  \city{Burnaby}
  \country{Canada}
}
\email{hza214@sfu.ca}

\author{Haisheng Fu}
\affiliation{%
  \institution{University of British Columbia}
  \city{Vancouver}
  \country{Canada}
}
\email{Haisheng.fu@ubc.ca}

\author{Zhengjie Miao}
\affiliation{%
  \institution{Simon Fraser University}
  \city{Burnaby}
  \country{Canada}
}
\email{zhengjie@sfu.ca}

\author{Zhenman Fang}
\affiliation{%
  \institution{University of Minnesota}
  \city{Twin Cities}
  \country{USA}
}
\email{zhenman@umn.edu}

\renewcommand{\shortauthors}{}

\begin{abstract}
Large language models (LLMs) enable semantic queries over unstructured data, but invoking an LLM on every row is prohibitively expensive. Proxy-based execution can reduce this cost by labeling a small sample with the LLM and training a lightweight proxy to label the remaining rows. Nevertheless, in practical scenarios where semantic queries are highly selective, it remains challenging to identify representative samples for labeling under a limited LLM budget and to achieve accurate predictions over the remaining data using the proxy.

We present PROBE, a learned proxy-based execution framework for highly selective multimodal semantic queries. PROBE uses selectivity-adaptive sampling to discover rare positive regions and trains a lightweight proxy on the acquired labels. It then refines proxy predictions through residual correction and homogeneity smoothing over a similarity graph. To address sampling bias, PROBE jointly calibrates refined scores using training labels and a separate calibration sample, then selectively routes candidates to the LLM by optimizing the tradeoff between estimated $F_1$ and evaluation cost. On queries from SemBench and LRO-Bench with positive rates as low as $2.17{\times}10^{-5}$, PROBE with a $10\%$ sampling budget yields a geometric-mean $F_1$ of $0.762$, compared with $0.797$ for exhaustive LLM evaluation, at a geometric-mean speedup of 4.87$\times$ and a maximum speedup of 9.3$\times$.

\end{abstract}

\maketitle

\pagestyle{plain}
\begingroup\small\noindent\raggedright\textbf{Artifact Availability:}\\
The source code, data, and/or other artifacts have been made available at \url{https://github.com/DamonZhao-sfu/PROBE}.
\par\endgroup

\section{Introduction}
\label{sec:intro}

Semantic query processing extends relational analytics with natural-language predicates over unstructured data, enabling semantic filters and joins beyond conventional SQL predicates~\cite{uqe,thalamusdb,liu2025palimpzest,patel2025lotus,cortex-aisql,bigquery-ai,databricks-ai}.
For example, the following query, simplified from MMQA Q7 in SemBench~\cite{lao2026sembench}, identifies airline-image pairs in which the image shows the airline's logo. Exhaustive execution evaluates the semantic predicate with a large language model (LLM) for every candidate pair, making the query prohibitively expensive even when only a few pairs match.

\begin{lstlisting}[
  language=SQL,
  basicstyle=\ttfamily\footnotesize,
  columns=fullflexible,
  keepspaces=true,
  breaklines=true,
  aboveskip=6pt,
  belowskip=6pt
]
  SELECT t.Airlines, i.uri
  FROM mmqa.airport t, mmqa.images i
  WHERE SEM_FILTER(
    "Determine if the image shows the logo of the airline.",
  t.Airlines, i.uri)
\end{lstlisting}

\begin{figure*}[t]
\centering
\includegraphics[width=0.9\textwidth]{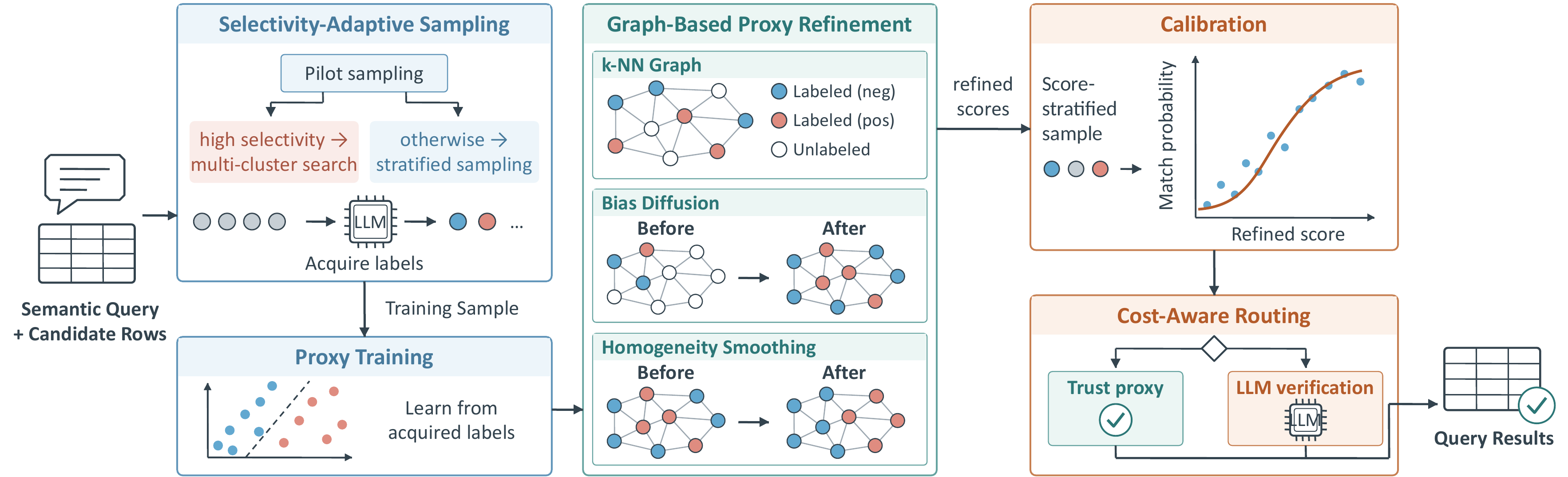}
\caption{Overview of PROBE: Selectivity-Adaptive Sampling and proxy training, graph-based proxy refinement, and Cost-Aware Routing between the proxy and the LLM.}
\label{fig:workflow}
\end{figure*}

A promising alternative is proxy-based execution~\cite{uqe,chung-proxy,zhang2026scaledoc,hou2026csv}: evaluate a
small sample with the LLM, train a lightweight classifier on the resulting labels, and use the classifier to process the remaining candidates.
However, this approach becomes unreliable for highly selective queries, where matching candidates are extremely rare. For example, MMQA Q7 contains only five matching pairs among 230,400 candidates, a positive rate of only $2.17\times 10^{-5}$. In such an extremely highly selective scenario, even uniformly sampling 10\% of the candidates has approximately a 59\% chance of missing all five
matches.\footnote{The probability is $\binom{207360}{5}/\binom{230400}{5}\approx 0.59$.}
Without positive examples, a proxy cannot learn to recognize the very results the query seeks. 
Highly selective semantic queries also arise in established benchmarks such as SemBench and LRO-Bench~\cite{lao2026sembench,lro}, as shown in Table~\ref{tab:workloads}, where the positive rates range from $2.17\times10^{-5}$ to $1.39\times10^{-1}$.

The challenge extends beyond discovering the first few matches. Rare positives may span across multiple separated regions of the input space, so sampling around discovered matches can leave other regions unexplored. 
Even when some positives are found, the limited and potentially biased training labels may not adequately support predictions on the remaining candidates.
Moreover, the scarcity of positives makes it difficult to estimate which proxy predictions are incorrect: a classifier can achieve high overall accuracy while missing many matches~\cite{bargain}. 
Re-evaluating uncertain predictions with the LLM can recover prediction quality, but undermines the cost savings that motivate proxy execution.
The central challenge is how to acquire and use limited LLM supervision effectively---to discover rare matches, generalize from scarce labels, and selectively correct remaining prediction errors.

\smallskip
\noindent\textbf{Limitations of Existing Systems.} Existing proxy-based systems generally follow two approaches: limiting the number of LLM evaluations through a fixed labeling
budget, or selecting proxy thresholds to satisfy a quality target.
Both face difficulties under high selectivity, but for different reasons.

\begin{figure}[t]
\centering
\includegraphics[width=\columnwidth]{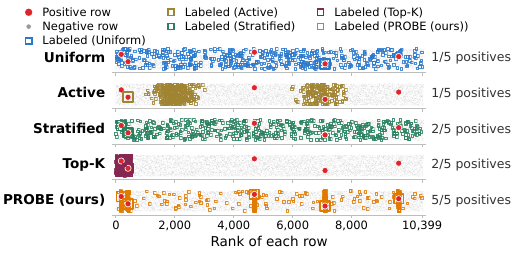}
\caption{Sampling outcomes on MMQA Q2a, with five matching pairs among 10.4k candidates ranked by descending semantic relevance score. Red dots indicate ground-truth matches, while squares indicate candidates selected for LLM labeling. With the same 5\% labeling budget, different sampling policies miss matches scattered across the ranking.}

\label{fig:sampler-miss}
\Description{Five horizontal strips show sampled rows by rank for uniform, active, stratified, top-K and PROBE sampling on a constructed ranking with five positives. The number of positives selected is shown beside each strip.}
\vspace{-5pt}
\end{figure}

\noindent\emph{Budget-based systems may miss rare matches.}
UQE~\cite{uqe} and CSV~\cite{hou2026csv} acquire a limited number of LLM labels for proxy-based execution.
When positives are extremely scarce, uniform or active sampling may miss entire regions containing matches, as illustrated in \autoref{fig:sampler-miss}. A proxy trained on these incomplete labels may classify the remaining matches as non-matches. In CSV, clusters with no positive samples can receive confident negative votes and bypass further re-clustering, leaving their matches undiscovered.

\noindent\emph{Target-based systems struggle to estimate quality reliably.}
LOTUS~\cite{patel2025lotus}, ScaleDoc~\cite{zhang2026scaledoc}, and BARGAIN~\cite{bargain} use labeled samples to select proxy thresholds under quality constraints. 
With very few positive examples, estimating precision or recall becomes difficult, potentially leading to conservative thresholds that require many LLM evaluations or poorly estimated thresholds that fail to meet the desired quality. Moreover, accuracy targets can be misleading under severe class imbalance: predicting every candidate as a non-match can achieve high accuracy while missing all matches, as illustrated by BARGAIN's accuracy mode.

These limitations reveal three interdependent challenges: acquiring labels that cover rare positive regions, leveraging scarce labels to generalize to unlabeled candidates, and identifying which remaining predictions require additional LLM evaluations. Importantly, the targeted sampling used to discover rare matches may also introduce selection bias, complicating the estimation of prediction errors and decisions about which predictions to verify.

To address these challenges, we present PROBE, a learned \underline{pro}xy-\allowbreak\underline{b}ased \underline{e}xe\-cu\-tion framework for highly selective multimodal semantic queries (\autoref{fig:workflow}). 
Given an initial sampling budget $B$ and a quality--cost tradeoff parameter $\alpha$, PROBE integrates targeted label acquisition, graph-based prediction refinement, and calibrated LLM routing for re-evaluation. Importantly, PROBE separates the training sample, which is selected to discover rare matches, from a calibration sample with known inclusion probabilities, enabling more reliable estimation of match probabilities among the remaining candidates. Our contributions are as follows:

\begin{enumerate}[leftmargin=*,itemsep=2pt,topsep=2pt,label=\arabic*)]

\item \emph{Selectivity-Adaptive Sampling (SAS).}
We propose an adaptive sampling strategy that uses a pilot sample to choose between stratified sampling and multi-cluster search.
For highly selective queries, SAS locates positive regions along the candidate ranking, expands sampling around discovered positive seeds and cluster boundaries, and reserves part of the labeling budget for uniform exploration.
This helps acquire rare positives and nearby negatives with limited LLM calls.

\item \emph{Graph-Based Proxy Refinement.}
We develop a graph-based refinement method that exploits the structure of unlabeled candidates to augment the
supervision provided by scarce LLM labels. 
It combines proxy residual correction and homogeneity smoothing over a query-conditioned similarity graph to refine scores for unlabeled inputs using the acquired LLM labels.

\item \emph{Learned Calibrated Cascade Routing.}
To address the sampling bias introduced by targeted label acquisition, PROBE fits a calibrator using the acquired training labels and a separate probability sample with known inclusion probabilities. It then uses calibrated match probabilities to select candidates for LLM re-evaluation, balancing their estimated contribution to $F_1$ against the additional evaluation cost.

\item \emph{Experimental evaluation.}
We evaluate PROBE on highly selective queries from SemBench and LRO-Bench, with positive rates as low as
$2.17\times 10^{-5}$. 
With a 10\% initial sampling budget, PROBE achieves a geometric-mean $F_1$ of $0.762$, compared with $0.797$ for exhaustive LLM evaluation, while achieving a geometric-mean speedup of $4.87\times$ and a maximum speedup of $9.3\times$.
\end{enumerate}

\section{Background}
\label{sec:bg}

\subsection{Semantic Operators}
\label{sec:bg:proxy}

A \emph{semantic filter} applies a natural-language predicate $\phi$ to each tuple of a relation $T$.
Let $y_\phi^{\mathrm{LLM}}(t)\in\{0,1\}$ denote the label returned by the LLM for tuple $t$.
\condr{Under exhaustive LLM evaluation, the filter returns $\mathrm{SemFilter}_\phi(T)=\{t\in T : y_\phi^{\mathrm{LLM}}(t)=1\}$.}

A \emph{semantic join} between relations $T^L$ and $T^R$ evaluates a natural-language predicate $\psi$ on tuple pairs.
\condr{Its output under exhaustive LLM evaluation is $\mathrm{SemJoin}_\psi(T^L,T^R)=\{(t^L,t^R)\in T^L\times T^R : y_\psi^{\mathrm{LLM}}(t^L,t^R)=1\}$.}

For both operators, let $\mathcal X=\{x_1,\ldots,x_N\}$ denote the set of candidate inputs,
where $x_i$ is a tuple for a semantic filter or a tuple pair for a semantic join. 
We use $y_i^{\mathrm{LLM}}\in\{0,1\}$ to denote the label returned by the LLM for candidate $x_i$.
These LLM labels are not assumed to be ground truth; benchmark ground-truth labels are used only for query result quality evaluation.

\subsection{Technical Background}
\label{sec:bg:technical}
\label{sec:bg:active}
\paragraph{Sampling.}
Uniform sampling selects candidates with equal probability, whereas importance sampling favors candidates according to
a cheap signal and uses sampling weights to estimate population quantities~\cite{supg,bas}.
Stratified sampling partitions candidates into disjoint groups and allocates samples across them.
Neyman allocation assigns samples in proportion to stratum sizes and within-stratum standard deviations to reduce
estimation variance~\cite{neyman1934,bas}. 
Under low positive rates, uniform sampling may yield few or no positive examples. The efficiency of importance sampling depends on whether the guiding signal identifies likely positives and informative candidates, while stratified allocation requires reliable estimates of within-stratum variances, which are difficult to obtain
when positive observations are scarce.

\label{sec:bg:ssl}
\paragraph{Semi-supervised Learning.}
Semi-supervised learning uses labeled and unlabeled data to improve predictions~\cite{zhu2005survey}. 
Graph-based methods construct a similarity graph over the inputs and propagate information from labeled to unlabeled nodes~\cite{zhu2003,zhou2003learning}. These methods typically assume local label consistency: connected nodes tend to have similar labels. 
Under low positive rates, positive labels may be too scarce to provide supervision for many regions containing matches, while propagation from negative labels may reinforce the majority class.
Moreover, when the similarity graph does not reflect label consistency, propagating labels can amplify errors~\cite{calder2020poisson, zhu2020beyond}.

\label{sec:bg:cascade}
\paragraph{Learned Proxy-Expert Routing.}
Proxy-expert routing decides whether an input is handled by a lightweight proxy or delegated to a more expensive
expert~\cite{routellm,hybridllm,posthoc-defer,gatekeeper}.
Such cascades have been studied in video analytics~\cite{noscope,blazeit,focus,tahoma} and semantic query processing~\cite{patel2025lotus,liu2025palimpzest, russo2025abacus,evaporate}.
Learned routing methods use expert-labeled examples and aim to estimate the reliability of proxy predictions, for example, by calibrating proxy scores into probabilities or learning a routing policy. 
These estimates guide delegation decisions to satisfy quality constraints or balance output quality against expert-evaluation cost ~\cite{liskowski2026streaming}.
Under low positive rates, scarce positive labels make it difficult to estimate proxy errors reliably. Overestimating these errors leads to excessive LLM calls, while underestimating them risks missing rare matches. Accuracy-based objectives may further reinforce such errors, as predicting all candidates as negative can still achieve high accuracy.

\subsection{Problem Definition}
\label{sec:bg:problem}
\label{def:problem}

Consider the $N$ candidates in $\mathcal{X} = \{x_1,\ldots,x_N\}$ of a semantic filter or join, with LLM labels $y_i^{\mathrm{LLM}}$ as defined in Section~\ref{sec:bg:proxy}.
Before routing, proxy-based execution acquires LLM labels for a set $S\subseteq \mathcal{X}$ of at most $B$ candidates for LLM evaluation and uses the acquired labels to train a proxy model and, when applicable, calibrate its prediction.
Let $\hat y_i^{\mathrm{proxy}}$ denote the proxy's predicted label for candidate $x_i$.
An additional set $D\subseteq \mathcal{X} \setminus S$ may then be sent to the LLM.
The final output for each candidate is
\[
\hat y_i(D)=
\begin{cases}
y_i^{\mathrm{LLM}}, & x_i\in S\cup D,\\
\hat y_i^{\mathrm{proxy}}, & \text{otherwise}.
\end{cases}
\]

Let $\hat{\mathbf y}(D)$ denote the resulting label vector over all $N$ candidates, and $\mathbf y^{\mathrm{LLM}}$ the corresponding vector of exhaustive LLM labels.
We measure execution quality by the $F_1$ score of $\hat{\mathbf y}(D)$ relative to $\mathbf y^{\mathrm{LLM}}$, and execution cost by the \emph{delegation rate}
$d(D)=\frac{|S|+|D|}{N}$,
the fraction of candidates evaluated by the LLM.

Given an initial sampling budget $B$ and a user-specified tradeoff parameter $\alpha\in[0,1]$, the objective is to minimize
\begin{equation}
\label{eq:problem}
\alpha\bigl(1-F_1(\hat{\mathbf y}(D), \mathbf y^{\mathrm{LLM}})\bigr) +(1-\alpha)d(D),
\end{equation}
subject to $|S|\leq B$. Larger $\alpha$ prioritizes output quality, while smaller $\alpha$ prioritizes reducing LLM evaluation cost. Since exhaustive LLM labels are unavailable during execution, the $F_1$ term must be estimated to guide routing decisions.

\section{Selectivity-Adaptive Sampling and Proxy Training}
\label{sec:sampling}

\subsection{Challenges and Design Goals}
\label{sec:sampling:problem}
\label{sec:sampling:reduction}

This section constructs the labeled set $L$ used to train the proxy. 
The initial sample $S$ defined in Section~\ref{sec:bg:problem} consists of the training set $L$ and a separate calibration sample $A$ (Section~\ref{sec:calib:model}), acquired at different stages.
PROBE splits the sampling budget $B$ into $B_L=\lfloor\kappa_L B\rfloor$ for proxy training and $B_A=B-B_L$ for calibration, where $\kappa_L=0.6$ by default is the fraction allocated to training.
Additional LLM calls from routing or whole-query fallback are not included in $B$, so the delegation rate may exceed $B/N$.

\noindent\textbf{Challenges.}
To train an accurate proxy, $L$ must contain sufficient positive examples and informative negatives near the decision boundary to distinguish matches from similar non-matches.
Under high selectivity and a small $B_L$, uniform sampling may find no positives, while top-$K$ sampling can miss positive regions below its cutoff.
Stratified sampling distributes the labeling budget across strata, but under high selectivity, even strata containing matches may yield no positive samples, limiting its ability to discover rare positives.
Similarly, UQE's active sampling~\cite{uqe} may fail to acquire positive labels when its initial sample contains no positives.

\noindent\textbf{Design goals.}
SAS therefore aims to (1) assess query selectivity with a small pilot sample, (2) find initial positive examples quickly, (3) explore around discovered positives to locate cluster boundaries and informative negatives, while reserving uniform samples to discover unseen positive clusters, 
and (4) train the proxy only when $L$ contains both of positive and negative class, falling back to exhaustive LLM evaluation otherwise.

\begin{figure}[t]
\centering
\resizebox{0.90\columnwidth}{!}{\begin{tikzpicture}[
  font=\normalsize,
  sfbox/.style={draw=black!70, rounded corners=2pt, align=center, inner sep=2.5pt, minimum height=5.5mm, fill=white},
  sfgate/.style={draw=black!70, diamond, aspect=2.8, align=center, inner sep=0.5pt, fill=black!4},
  sfend/.style={draw=black!70, rounded corners=2pt, align=center, inner sep=2.5pt, minimum height=5.5mm, fill=black!8},
  sfarrow/.style={-{Stealth[length=1.6mm]}, draw=black!75, line width=0.5pt},
  sflab/.style={text=black!80},
]
\node[sfbox, text width=7.6cm] (pilot) at (3.9,0.3) {Pilot sampling on rows sorted by encoder score\\$\rightarrow$ $95\%$ lower bound $\pi_{\mathrm{low}}$ on the positive rate};
\node[sfgate] (gate) at (3.9,-0.95) {$\pi_{\mathrm{low}}>p_\star$?};

\node[sfbox, text width=3.9cm] (explore) at (2.3,-1.95) {\textbf{Explore phase}\\scan from the highest scores};
\node[sfbox, text width=3.9cm] (exploit) at (2.3,-3.15) {\textbf{Exploit phase}\\fit clusters and windows};
\node[sflab, text width=3.9cm, align=center] (note) at (2.3,-3.85) {+ uniform exploration};
\begin{scope}[on background layer]
\node[draw=black!45, dashed, rounded corners=3pt, fit=(explore)(exploit)(note), inner xsep=3pt, inner ysep=2pt,
      label={[sflab, rotate=90, anchor=south, inner sep=1pt]west:Multi-cluster search}] (multi) {};
\end{scope}

\node[sfbox, text width=2.9cm] (strat) at (6.35,-2.5) {\textbf{Stratified}\\\textbf{sampling}};

\coordinate (sfmerge) at (3.9,-4.4);
\node[sfgate] (both) at (3.9,-5.55) {Has both classes?};
\node[sfend, text width=1.75cm] (train) at (0.35,-5.55) {Train proxy\\(\S\ref{sec:sampling:proxy})};
\node[sfend, text width=1.75cm] (fallback) at (7.45,-5.55) {LLM on\\all rows};

\draw[sfarrow] (pilot) -- (gate);
\draw[sfarrow] (gate.west) -| (explore.north);
\node[sflab, anchor=south east, align=right, inner sep=1pt] at ([xshift=-3pt]gate.west) {no\\{\small (highly selective)}};
\draw[sfarrow] (gate.east) -| (strat.north);
\node[sflab, anchor=south west, align=left, inner sep=1pt] at ([xshift=3pt]gate.east) {yes\\{\small (not highly selective)}};
\draw[sfarrow] (explore) -- node[sflab, right] {$n_+\ge n_{\min}$} (exploit);
\draw[draw=black!75, line width=0.5pt] (multi.south) |- (sfmerge);
\draw[draw=black!75, line width=0.5pt] (strat.south) |- (sfmerge);
\draw[sfarrow] (sfmerge) -- (both.north);
\draw[sfarrow] (both.west) -- node[sflab, above] {yes} (train.east);
\draw[sfarrow] (both.east) -- node[sflab, above] {no} (fallback.west);

\end{tikzpicture}}
\caption{Selectivity-Adaptive Sampling. Pilot sampling selects between stratified sampling and multi-cluster search
based on estimated selectivity.}
\label{fig:sampling-stage}
\end{figure}
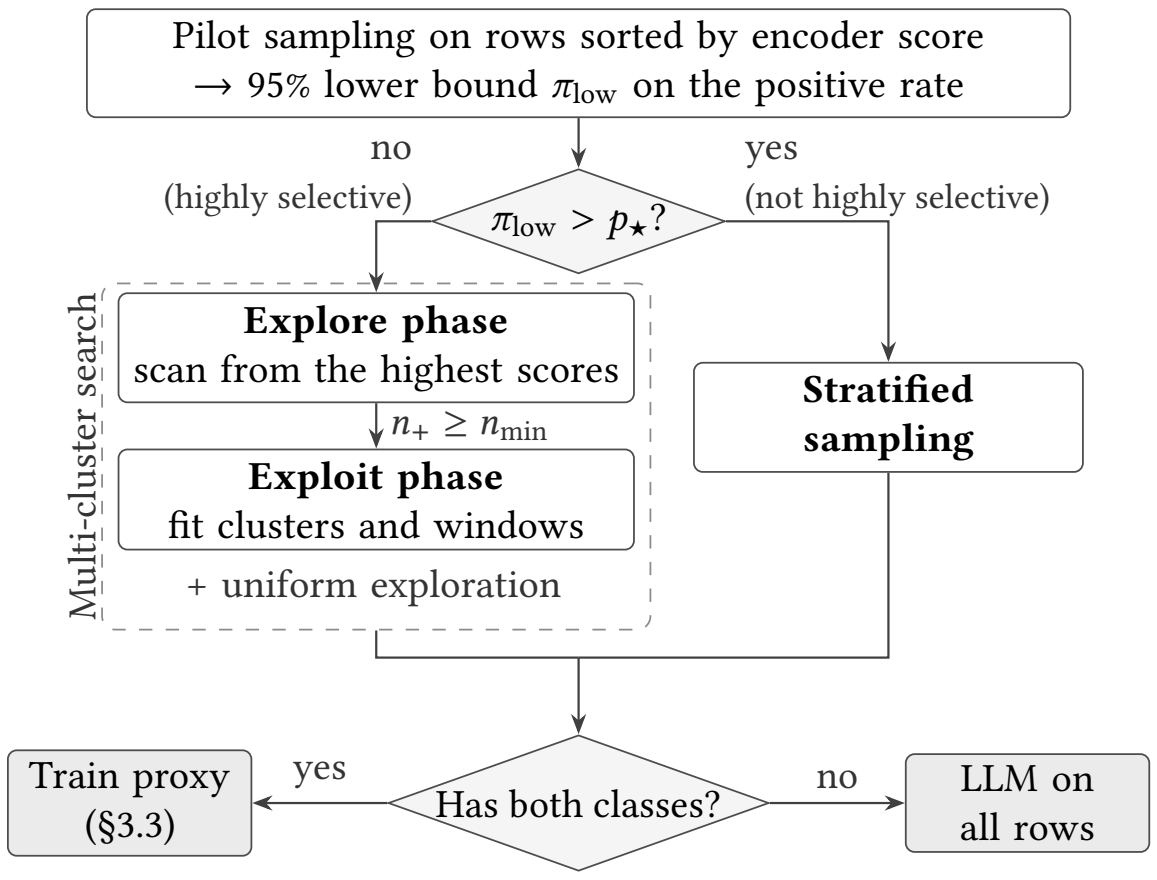

\subsection{Selectivity-Adaptive Sampling}
\label{sec:sampling:algo}

\textbf{Overview.} 
Selectivity-Adaptive Sampling (SAS) processes candidates ranked by descending cross-encoder score $s_i$, computed in one pass that reads the predicate and the candidate together without LLM evaluation. 
As shown in Algorithm~\ref{alg:active-sampling}, it first labels a pilot sample and computes a lower confidence bound on the positive rate.
If the bound exceeds the threshold $p_\star$, SAS uses stratified sampling; otherwise, it performs the multi-cluster search of Algorithm~\ref{alg:multi-cluster}.
Before sufficient positives are discovered to form clusters, the search prioritizes the highest-scoring candidates.
During multi-cluster search, each batch also reserves a share for uniform sampling across the ranking, allowing SAS to find positives that lie far below the highest-scoring candidates.
SAS returns $L$ for proxy training if it contains both positive and negative class, and otherwise falls back to exhaustive LLM evaluation.

\begin{algorithm}[t]
\small
\caption{Selectivity-Adaptive Sampling (SAS)}
\label{alg:active-sampling}
\label{alg:rts}
\begin{algorithmic}[1]
\Statex \textbf{Input:} ranked rows $\mathcal X$, budget $B_L$, SAS parameters (\S\ref{sec:sampling:algo}).
\State $X_0\gets\Call{SampleUniform}{\mathcal X,\lfloor f_{\mathrm{pilot}}B_L\rfloor}$

\State $L\gets\Call{Label}{X_0}$,\ $\pi_{\mathrm{low}}\gets\text{Eq.~\eqref{eq:gate}}$
\If{$\pi_{\mathrm{low}}>p_\star$}
\State $L\gets\Call{Stratified}{L}$
\Else
\State $L\gets\Call{MultiCluster}{L}$
\EndIf
\If{$L$ does not include both positive and negative classes}
\State \Return {$L\cup\Call{Label}{\mathcal X\setminus L}$}\Comment{{whole query to the LLM}}
\EndIf
\State \Return {$L$}
\Statex

\Function{Stratified}{$L$}
\State $H\gets\min(M_{\mathrm{train}},\lfloor\log_2N\rfloor+1)$
\State $w_0\gets N/(2^H-1)$
\For{$h=1,\ldots,H$}
\State $\mathcal X_h\gets\{i:w_0(2^{h-1}-1)\le r_i<w_0(2^h-1)\}$
\State $\hat\pi_h\gets\dfrac{\sum_{i\in X_0\cap\mathcal X_h}y_i^{\mathrm{LLM}}}{\max(1,|X_0\cap\mathcal X_h|)}$
\State $\eta_h\gets|\mathcal X_h|\sqrt{\hat\pi_h(1-\hat\pi_h)}$
\EndFor
\While{$|L|<\min(B_L,N)$}
\State $U\gets\mathcal X\setminus L$,\ $b\gets\min(b_0,B_L-|L|,|U|)$

\State $(b_h)_{h=1}^H\gets\Call{NeymanAlloc}{b,(\eta_h)_h,(|\mathcal X_h\cap U|)_h}$
\State $L\gets L\cup\Call{Label}{\bigcup_{h=1}^H\textsc{SampleUniform}(\mathcal X_h\cap U,b_h)}$

\EndWhile
\State \Return $L$

\EndFunction
\end{algorithmic}
\end{algorithm}

\begin{algorithm}[t]
\small
\caption{Multi-cluster search}
\label{alg:multi-cluster}
\begin{algorithmic}[1]
\Statex \textbf{Input:} pilot-labeled $L$, ranked rows $\mathcal X$, budget $B_L$, batch $b_0$, exploration share $\varepsilon_{\mathrm{explore}}$, explore threshold $n_{\min}$, cluster penalty $\mu$, maximum cluster counts $I_{\max}$.
\Function{MultiCluster}{$L$}
\State $\lambda\gets0$,\ $j\gets0$,\ $\mathcal W\gets\emptyset$
\While{$|L|<\min(B_L,N)$}
\State $U\gets\mathcal X\setminus L$,\ $b\gets\min(b_0,B_L-|L|,|U|)$ , $\lambda\gets\lambda+\varepsilon_{\mathrm{explore}}b$

\State $T\gets\Call{SampleUniform}{U,\lfloor\lambda\rfloor}$,\ $\lambda\gets\lambda-|T|$ \Comment{Explore}
\If{$n_+<n_{\min}$}
\State $T\gets T\cup\Call{SampleHighestScore}{U\setminus T,b-|T|}$
\Else\Comment{Exploit}
\State $j\gets j+1$;\ \textbf{if} $j=1$: $\rho\gets N^{1/\lceil(B_L-|L|)/b_0\rceil}$
\State $\mathcal I\gets\Call{FitClusters}{L}$\Comment{Eq.~\eqref{eq:clusters}}

\State $\mathcal S\gets\{i\in L:y_i^{\mathrm{LLM}}=1,\ r_i\notin I\ \forall I\in\mathcal I\}$\Comment{seeds}

\State $\mathcal G\gets\Call{Gaps}{\mathcal I,\mathcal S,L}$\Comment{$g\in\mathcal G$: $|g|$ rows, center rank $c_g$}
\State $w_g\gets$ {width of the window in $\mathcal W$ with $c_g$}, else $N/\rho^{\,j-1}$
\State $w_g\gets\min\bigl(N,\max(1,|g|,\lceil w_g/\rho\rceil)\bigr)$\Comment{$\forall g\in\mathcal G$}
\State $\mathcal W\gets\Call{Merge}{\mathcal I\cup\{\textsc{Window}(c_g,w_g)\}_{g\in\mathcal G}}$
\State $T\gets T\cup\Call{SampleNearSeeds}{U\setminus T,\mathcal S,\lfloor(b-|T|)/2\rfloor}$
\State $T\gets T\cup\Call{SampleInWindow}{U\setminus T,\mathcal W,b-|T|}$
\EndIf
\State $L\gets L\cup\Call{Label}{T}$
\EndWhile
\State \Return $L$
\EndFunction
\end{algorithmic}
\end{algorithm}

\noindent\textbf{Notation and parameters}. Let $r_i\in\{0,\ldots,N-1\}$ denote the rank of candidate $x_i$ in descending score order, 
and let $U=\mathcal X\setminus L$ denote the unlabeled candidates.
The current class counts are $n_+=\sum_{x_i\in L}y_i^{\mathrm{LLM}}$ and $n_-=|L|-n_+$.
For a candidate set $T$, $\textsc{SampleUniform}(T,m)$ draws $\min(m,|T|)$ candidates uniformly,
$\textsc{SampleHighestScore}(T,m)$ selects the $\min(m,|T|)$ highest-scoring candidates, and $\textsc{Label}(T)$ contains their LLM labels.
The default parameters are the batch size $b_0=\max(1,\lfloor0.01N\rfloor)$, pilot share $f_{\mathrm{pilot}}=0.3$, selectivity threshold $p_\star=0.05$, stratum cap $M_{\mathrm{train}}=10$, Explore threshold $n_{\min}=2$, exploration share $\varepsilon_{\mathrm{explore}}=0.25$, cluster penalty $\mu=1.5$, and maximum cluster counts $I_{\max}=6$.

\textbf{Pilot sampling.}
We call a query highly selective if its LLM-positive rate is below $p_\star=0.05$.
SAS draws a uniform pilot sample $X_0$ of $n_0$ candidates, of which $n_0^+$ are positive.
It computes a one-sided $95\%$ lower confidence bound on the positive rate by inverting the hypergeometric upper tail:
\begin{equation}
\label{eq:gate}
\pi_{\mathrm{low}}= \min\bigl\{K/N:\ n_0^+\le K\le N,\ \Pr[X_K\ge n_0^+]>0.05\bigr\},
\end{equation}
where $X_K\sim\operatorname{Hypergeom}(N,K,n_0)$ is the number of positives in a pilot sample drawn from a population with $K$ positives
If $\pi_{\mathrm{low}}>p_\star$, SAS rules out high selectivity at the $95\%$ confidence level and uses stratified sampling. Otherwise, SAS cannot rule out high selectivity and starts a multi-cluster search.

\noindent\textbf{Stratified sampling.} 
When the pilot rules out high selectivity, it already contains positive examples, so SAS focuses the remaining budget on acquiring informative training labels. \textsc{Stratified} is described in Algorithm~\ref{alg:active-sampling}.
SAS divides the score ranking into $H$ strata whose widths double with rank and estimates each stratum's positive rate $\hat\pi_h$ from the pilot sample.
It assigns stratum $h$ the Neyman weight $\eta_h=|\mathcal X_h|\sqrt{\hat\pi_h(1-\hat\pi_h)}$~\cite{neyman1934}.
Large strata with mixed pilot labels receive higher weights, while strata with only one observed class receive zero weight.
For each batch of $b$ labels, \textsc{NeymanAlloc} allocates samples across strata according to these weights (i.e., $b_h= b\,\eta_h/\sum_j\eta_j$), with minimum allocations to maintain coverage. 
SAS samples uniformly within each stratum.

\noindent\textbf{Multi-cluster search.}
When pilot sampling cannot rule out high selectivity, SAS searches for rare positives and their neighboring
negatives along the score ranking.
It models positive regions as at most $I_{\max}$ contiguous clusters, treating candidates outside these
clusters as negative during fitting.
A \emph{cluster} is a contiguous rank range supported by labeled positives, while a labeled positive outside every fitted cluster is a \emph{seed}. The rank ranges around cluster boundaries and seeds form \emph{windows} for subsequent sampling.

\begin{figure}[t]
\centering
\includegraphics[width=\linewidth]{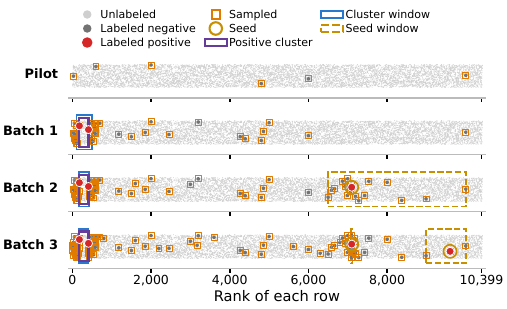}
\vspace{-5mm}
\caption{Multi-cluster search on MMQA Q2a among 10.4k candidates. Blue and orange boxes show the selected windows used for sampling; purple boxes show the positive clusters. Red points show selected positive samples, and orange circles mark selected seed samples.}
\label{fig:sas-multi-process}
\vspace{-3mm}
\end{figure}

While $L$ contains fewer than $n_{\min}$ positives, SAS is in the \emph{Explore} phase: with too few positives to fit clusters, it prioritizes the highest-ranked unlabeled candidates to find initial positives.
In every batch, a share $\varepsilon_{\mathrm{explore}}$ of the labeling budget is reserved for uniform sampling, allowing SAS to discover positives far below the top-ranked candidates.
Once enough positives are found, SAS enters the \emph{Exploit} phase: it refits the cluster model of~\autoref{eq:clusters} using all acquired labels and directs sampling toward seeds and windows around fitted cluster boundaries, while continuing uniform exploration. 
As labels accumulate, clusters may emerge, grow, shrink, split, or merge. \autoref{fig:sas-multi-process} illustrates the process.

\noindent\textbf{Fitting clusters}. \textsc{FitClusters} covers the labeled positive samples with at most $I_{\max}$ disjoint clusters $\mathcal I$. For a set of clusters $\mathcal I$, let $n^+_{\mathrm{out}}(\mathcal I)$ be the number of labeled positive samples outside every cluster, and $n^-_{\mathrm{in}}(\mathcal I)$ the number of labeled negative samples inside a cluster. The clusters minimize the sum of three terms,
\begin{equation}
\label{eq:clusters}
\underbrace{n^+_{\mathrm{out}}(\mathcal I)}_{\text{\scriptsize positive samples outside}}
+\underbrace{n^-_{\mathrm{in}}(\mathcal I)}_{\text{\scriptsize negative samples inside}}
+\underbrace{\mu|\mathcal I|}_{\text{\scriptsize penalty per cluster}},
\end{equation}
where $|\mathcal I|$ is the number of clusters. PROBE minimizes Eq.~\eqref{eq:clusters} with the dynamic program in Eq.~\eqref{eq:dp-recurrence}. Eq.~\eqref{eq:dp-recurrence} follows the standard dynamic program for optimally partitioning ordered data with a penalty per segment~\cite{bellman1961approximation,jackson2005algorithm}. Unlike the standard form, segments alternate between inside and outside; only inside segments pay $\mu$, and their number is capped at $I_{\max}$. Eq.~\eqref{eq:dp-recurrence} scans the labeled rows in rank order and adds the terms of Eq.~\eqref{eq:clusters} one row at a time. Let $y_i\in\{0,1\}$ be the LLM label of the $i$-th labeled row, $\mathrm{In}_i[q]$ be the lowest value of Eq.~\eqref{eq:clusters} over the first $i$ labeled rows using $q$ clusters when row $i$ lies inside the $q$-th cluster, and let $\mathrm{Out}_i[q]$ be the same when row $i$ lies outside every cluster. Two states are needed because the cost of putting a row inside a cluster depends on the previous row. If the previous row is inside a cluster, the row extends that cluster at no cost. If the previous row is outside, the row starts a new cluster and adds $\mu$. Then
\begin{equation}
\begin{aligned}
\mathrm{In}_i[q]  &= (1-y_i) + \min\bigl(\mathrm{In}_{i-1}[q],\ \mathrm{Out}_{i-1}[q-1]+\mu\bigr),\\
\mathrm{Out}_i[q] &= y_i + \min\bigl(\mathrm{In}_{i-1}[q],\ \mathrm{Out}_{i-1}[q]\bigr),
\end{aligned}
\label{eq:dp-recurrence}
\end{equation}
for $1\le i\le |L|$, with $1\leq q\le I_{\max}$ for $\mathrm{In}$ and $0\leq q\le I_{\max}$ for $\mathrm{Out}$. In Eq.~\eqref{eq:dp-recurrence}, the first term is the cost of row $i$ itself: $1-y_i$ counts a negative sample inside a cluster, and $y_i$ counts a positive sample outside every cluster. The minimum picks the cheaper state of row $i-1$. The smallest entry of $\mathrm{In}_{|L|}[q]$ and $\mathrm{Out}_{|L|}[q]$ over $0\leq q\le I_{\max}$ is the minimum of Eq.~\eqref{eq:clusters}, and backtracking recovers $\mathcal I$. The penalty $\mu$ sets how much labeled evidence a cluster needs. A cluster that covers $n_I^+$ labeled positives and $n_I^-$ labeled negatives lowers Eq.~\eqref{eq:clusters} only if $n_I^+-n_I^->\mu$. For example, with $\mu=1.5$, a cluster must cover at least two more positive than negative samples. A lone positive sample stays a seed, while two positives with no labeled negative between them can support a cluster.

\noindent\textbf{Windows and allocation.}
Let $\mathcal S$ be the set of seeds, i.e., labeled positives outside the fitted clusters $\mathcal I$.
The boundaries of these clusters and the extents of positive regions around seeds remain uncertain within the unlabeled \emph{gaps} up to the nearest labeled candidates.
$\textsc{Gaps}(\mathcal I,\mathcal S,L)$ identifies these gaps; let $|g|$ and $c_g$ denote the size and center rank of gap $g$.
Since early clusters may have inaccurate boundaries, SAS samples from wider \emph{windows} around these gaps, where $\textsc{Window}(c,w)$ denotes the $w$ ranks centered at $c$.
Let $\mathcal W$ be the windows from the previous Exploit batch.
In the $j$-th Exploit batch, a gap centered within a previous window inherits its width reduced by a factor of $\rho$; otherwise, its window starts with width $N/\rho^j$. 
SAS sets $\rho$ in the first Exploit batch so that $N/\rho^j$ approaches one when the sampling budget runs out. 
Fitted clusters are also included as windows, allowing negative labels within them to trigger splits, and $\textsc{Merge}$ combines overlapping windows.
After reserving the uniform exploration share, $\textsc{SampleNearSeeds}(U,\mathcal S,m)$ uses up to half of the remaining budget to sample near seeds in round-robin order, or sample uniformly if no seeds exist.
$\textsc{SampleInWindow}(U,\mathcal W,m)$ allocates the remaining budget across windows: each cluster-associated
window receives one share, while seed-only windows together receive one share. Candidates are sampled uniformly within each window.

\subsection{Proxy Training}
\label{sec:sampling:proxy}
PROBE trains a lightweight proxy on the LLM-labeled set $L$ acquired by SAS.
For each candidate $x_i$, the cross-encoder used for ranking processes the natural-language predicate together
with the candidate and produces a score $s_i$.
We also extract its last-token hidden state, normalize it to unit length, and project it to $256$ dimensions using PCA. 
The resulting feature vector $e_i$ represents $x_i$ conditioned on the predicate. For a semantic join, the cross-encoder processes both tuples of the candidate pair, so $e_i$ represents the pair rather than either tuple alone.

The proxy $\ell$ is an $L_2$-regularized logistic regression model trained on $L$, with weight vector $v$ and intercept $v_0$.
Its inverse regularization strength $C_{\mathrm{LR}}$ controls the penalty on model weights; larger values correspond to weaker regularization ($C_{\mathrm{LR}}=1$ by default).
To mitigate class imbalance, we weight each class inversely to its frequency in $L$.
Applying $\ell$ to all candidates produces proxy scores
\[
Z_i=\operatorname{sig}(v^\top e_i+v_0)\in(0,1),
\]
where $\operatorname{sig}$ is the logistic sigmoid function.

\section{Graph-Based Proxy Refinement}
\label{sec:ssl}

\begin{algorithm}[t]
\caption{Graph-Based Proxy Refinement}
\label{alg:label-prop}
\begin{algorithmic}[1]
\Statex \textbf{Input:} row embeddings $e$, proxy scores $Z$, labeled set $L$ with LLM labels $y_L^{\mathrm{LLM}}$, neighbor count $k$, diffusion and propagation rounds $T_1,T_2$, diffusion decay $\beta$, propagation weight $\gamma$, label threshold $\theta_{\mathrm{ref}}$.
\State $W\gets\Call{SymmetricKNN}{e,k}$ 
\State $P\gets\operatorname{diag}(W\mathbf1)^{-1}W$
\State $\Delta^{(0)}\gets\mathbf0$,\ $\Delta_L^{(0)}\gets y_L^{\mathrm{LLM}}-Z_L$
\For{$t=1,\ldots,T_1$} 
\State $\Delta^{(t)}\gets\beta P\Delta^{(t-1)}$
\State $\Delta_L^{(t)}\gets y_L^{\mathrm{LLM}}-Z_L$
\EndFor
\State $f^{(0)}\gets\operatorname{clip}(Z+\Delta^{(T_1)},0,1)$
\For{$t=1,\ldots,T_2$}
\State $f^{(t)}\gets(1-\gamma)f^{(0)}+\gamma Pf^{(t-1)}$
\State $f_L^{(t)}\gets y_L^{\mathrm{LLM}}$
\EndFor
\State $f\gets f^{(T_2)}$
\State $y^{\mathrm{refined}}\gets\mathbbm1[f\ge\theta_{\mathrm{ref}}]$
\State \Return $(f,y^{\mathrm{refined}})$
\end{algorithmic}
\end{algorithm}

\subsection{Refinement Assumption}
\label{sec:ssl:objective}

The proxy scores each row from its own feature, independently of the others. PROBE adds a weighted $k$-nearest-neighbor ($k$-NN) graph to share residual error and score information between nearby rows, using the LLM labels in $L$ as anchors. The refinement relies on two empirical assumptions. First, local
error correlation assumes that when the proxy over- or underestimates the score of one labeled row, nearby rows may exhibit a
residual error $y_i^{\mathrm{LLM}}-Z_i$ in the same direction. Second, local homophily assumes that nearby rows
tend to have the same LLM-predicted label. The two assumptions motivate propagating the residuals and smoothing the corrected scores, respectively~\cite{zhou2003learning,correct-smooth}.

\subsection{Graph Construction}
\label{sec:ssl:graph}

$\textsc{SymmetricKNN}$ in Algorithm~\ref{alg:label-prop} builds a symmetric $k$-NN graph whose nodes are the input rows (pairs for a join), each represented by its cross-encoder feature $e_i$. PROBE first normalizes each embedding, $\tilde e_i=e_i/\lVert e_i\rVert_2$, and connects every input to its $k$ nearest neighbors under Euclidean distance. It assigns each edge the Gaussian weight $W_{ij}=\exp(-\lVert\tilde e_i-\tilde e_j\rVert_2^2/2\sigma^2)$ for $j\in\mathcal N_i$ and sets $W_{ij}=0$ otherwise. The bandwidth $\sigma$ is the median distance
from each embedding to its $k$th nearest neighbor. $\mathcal N_i$ contains both the $k$ neighbors selected by row $i$ and any rows that select $i$ as one of their own nearest neighbors. Both refinement phases use $P=\operatorname{diag}(W\mathbf{1})^{-1}W$. Here, $\mathbf1$ is the all-ones vector, $\operatorname{diag}(z)$ denotes the diagonal matrix with diagonal $z$.
\subsection{Score Propagation}
\label{sec:ssl:prop}

\begin{figure}[t]
\centering
\includegraphics[width=0.80\columnwidth]{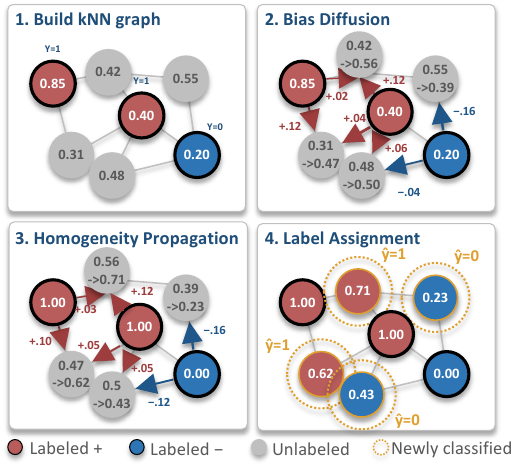}
\vspace{-5mm}
\caption{
Graph-based proxy refinement: \textbf{(1)} build a $k$-NN graph with LLM labels and proxy scores; \textbf{(2)} correct scores by bias diffusion; \textbf{(3)} smooth scores while pinning labeled rows; and \textbf{(4)} threshold at $\theta_{\mathrm{ref}}$ to assign refined labels.
}
\label{fig:label-prop-pipeline}
\vspace{-1mm}
\end{figure}

Score propagation returns refined scores $f_{i}\in[0,1]$ and refined labels ${y_i^{\mathrm{refined}}}\in\{0,1\}$, without retraining the proxy or calling the LLM. \autoref{fig:label-prop-pipeline} illustrates how the algorithm works.

\textbf{Bias diffusion.} PROBE initializes $\Delta_L^{(0)}=y_L^{\mathrm{LLM}}-Z_L$ and $\Delta_U^{(0)}=0$.
At each iteration, it applies $\beta P$ and restores the observed residuals on $L$.
After $T_1$ iterations, PROBE clips $Z+\Delta^{(T_1)}$ to $[0,1]$ and sets its labeled entries to $y_L^{\mathrm{LLM}}$.

\textbf{Homogeneity propagation.} At each iteration, PROBE combines the corrected vector $f^{(0)}$ and $Pf^{(t-1)}$ with weights $1-\gamma$ and $\gamma$, then restores the labels on $L$.
After $T_2$ iterations, it thresholds the unlabeled scores at $\theta_{\mathrm{ref}}$ ($0.5$ by default).

\section{Learned Calibrated Cascade Routing}
\label{sec:calib}
\label{sec:ssl:deferral}

Target-based cascades select candidates for LLM evaluation to meet a user-specified quality target. 
Under low positive rates, scarce positive labels make it difficult to assess whether a routing policy meets the target. PROBE instead estimates output quality from learned match probabilities and optimizes its tradeoff with LLM cost. 
Overestimating the benefit of LLM verification can lead to unnecessary LLM calls, while underestimating it can leave proxy errors uncorrected, including missed rare positives. Learning reliable match probabilities is challenging under high selectivity. SAS concentrates positive examples in $L$, but its preference for likely matches makes $L$ unrepresentative of the remaining candidates.
A calibrator fitted on $L$ alone may overestimate their match probabilities. 
In contrast, uniform sampling of the remaining candidates may yield too few positives for reliable calibration. Positive-label scarcity can also delay calibration in GAMCAL~\cite{liskowski2026streaming}, which samples uniformly from its uncertain region and requires examples of both classes.
PROBE therefore fits a calibrator jointly on $L$ and a separate calibration sample $A$. The positives in $L$
help learn the relationship between refined scores and LLM labels, while the known inclusion probabilities of $A$ support weighted calibration that corrects for the selection bias in $L$.

\begin{figure}[t]
\centering
\includegraphics[width=\columnwidth]{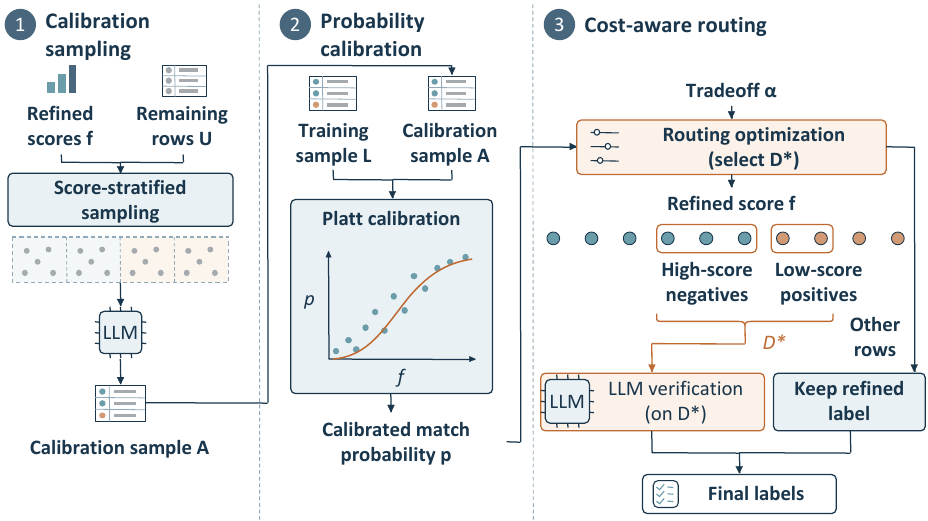}
\vspace{-5mm}
\caption{Learned calibration and cost-aware routing.}
\label{fig:calibration-routing}
\Description{Score-stratified calibration sampling, Platt calibration using training and calibration samples, and cost-aware routing to the LLM or refined labels.}
\end{figure}

\autoref{fig:calibration-routing} illustrates the three steps, summarized in Algorithm~\ref{alg:learned-calibration}.
First, PROBE draws a score-stratified calibration sample $A$ from $U=\mathcal X\setminus L$ and obtains its LLM labels.
Second, it fits the calibrator on $L$ and $A$ and converts the refined scores of candidates outside $S=L\cup A$ into match probabilities $p_i$.
Finally, guided by the quality--cost weight $\alpha$, it selects a set $D^*\subseteq\mathcal X\setminus S$ for LLM verification, potentially including high-score predicted negatives and low-score predicted positives. Other candidates retain their refined labels.
If $A$ or $\mathcal X\setminus S$ is empty, or calibration fails, PROBE keeps the available LLM labels and uses refined predictions for the rest.

\begin{algorithm}[t]
\small
\caption{Learned Calibrated Cascade Routing}
\label{alg:learned-calibration}
\begin{algorithmic}[1]
\Statex \textbf{Input:} rows $\mathcal X$, labeled set $L$ with LLM labels $y_L^{\mathrm{LLM}}$, refined scores $f$ and labels $y^{\mathrm{refined}}$, out-of-fold scores $f_L^{\mathrm{oof}}$, calibration budget $B_A$, score bins {$n_{\mathrm{bin}}$}, even-allocation share $\varepsilon_A$, quality--cost weight $\alpha$.
\State $U\gets\mathcal X\setminus L$,\ $m\gets\min(B_A,|U|)$\Comment{calibration budget}
\State $(A,\omega)\gets\Call{CalibrationSample}{U,f,m,{n_{\mathrm{bin}}},\varepsilon_A}$
\State $A\gets\Call{Label}{A}$,\ $S\gets L\cup A$
\State $\hat y\gets y^{\mathrm{refined}}$,\ $\hat y_S\gets y_S^{\mathrm{LLM}}$
\If{$A=\emptyset$ or $S=\mathcal X$}
\State \Return $\hat y$
\EndIf

\State $(a,c)\gets\Call{Calibrate}{A,L,\omega,f,f_L^{\mathrm{oof}}}$ 
\If{calibration fails}
\State \Return $\hat y$
\EndIf
\State $p_i\gets\operatorname{sig}(a{\operatorname{logit}_{\epsilon}}(f_i)+c),\quad i\notin S$

\State $D^*\gets\displaystyle\arg\min_{D\subseteq\mathcal X\setminus S}\bigl\{\alpha(1-\widetilde F_1(D))+(1-\alpha)\tfrac{|S|+|D|}{N}\bigr\}$
\State $D^*\gets\Call{Label}{D^*}$,\ $\hat y_{D^*}\gets y_{D^*}^{\mathrm{LLM}}$
\State \Return $\hat y$
\end{algorithmic}
\end{algorithm}

\subsection{Learned Calibration}
\label{sec:calib:model}

\textbf{Calibration Set Sampling.} The calibration sample $A$ provides labels from $U$ for the weighted fit. PROBE draws $m=\min(B_A,|U|)$ rows from $U$ with known inclusion probabilities, so each row of $A$ can be weighted by the number of rows of $U$ it represents.
A uniform sample would spend most draws on low-score rows that are almost all negative, so PROBE stratifies $U$ by refined score and uses score-based estimates of within-stratum variability to guide sample allocation.
It partitions $U$ into $n_{\mathrm{bin}}=10$ equal-width score bins and drops empty bins. While more than {$\lceil\varepsilon_A m\rceil$} bins remain, it merges the adjacent pair with the smallest combined size {, which leaves $H_A\le\lceil\varepsilon_A m\rceil$ strata $U_1,\ldots,U_{H_A}$}.
Stratum $U_j$ has size $N_j=|U_j|$, mean refined score $\bar f_j$, and Neyman weight {$\eta_j$}$=N_j\sqrt{\bar f_j(1-\bar f_j)}$. {PROBE spreads $\lceil\varepsilon_A m\rceil$ draws evenly over the strata, one at a time to a non-full stratum with the fewest draws, so every stratum gets at least one. It splits the other draws in proportion to $\eta_j$~\cite{neyman1934} over the non-full strata} and assigns leftover draws by the same fewest-draws rule. The final quotas satisfy $1\le n_j\le N_j$ and $\sum_{j=1}^{H_A}n_j=m$. $\textsc{CalibrationSample}$ draws $n_j$ rows uniformly from each stratum and returns $A$ with $\omega_i=N_j/n_j$ for $i\in A\cap U_j$.

\textbf{Calibrator.}
PROBE uses a Platt-style model~\cite{platt1999probabilistic}
to estimate LLM match probabilities from refined scores.
Eq.~\eqref{eq:platt} specifies the model, and Eq.~\eqref{eq:loss} gives its fitting objective.
At a finite optimum, the fitted probabilities satisfy Eq.~\eqref{eq:anchor}: the predicted and observed positive counts on $A$ agree after inverse-probability weighting.

\begin{equation}\label{eq:platt}
{g_i=\operatorname{sig}\big(a\,\operatorname{logit}_{\epsilon}(\hat f_i)+\hat c_i\big),\qquad a\ge 0,}
\end{equation}
{where $(\hat f_i,\hat c_i)=(f_i^{\mathrm{oof}},c_L)$ for $i\in L$ and $(f_i,c)$ for $i\in U$, $f_i$ is the refined score from Algorithm~\ref{alg:label-prop}, $\operatorname{sig}(x)=1/(1+e^{-x})$, and $\operatorname{logit}_{\epsilon}(f)=\ln\big(\bar f/(1-\bar f)\big)$ with $\bar f=\min(1-\epsilon,\max(\epsilon,f))$.
The slope $a$ sets how fast the match probability rises with the refined score, and the intercept sets the overall level of the probabilities, i.e., how many positive samples the model expects. 
The shared slope $a$ learns the score--label relationship from both samples, while separate intercepts $c_L$ and $c$ allow their baseline match probabilities to differ.
This parameterization is motivated by the intercept shift under case--control sampling~\cite{prentice1979logistic,king2001logistic}, although it remains a modeling approximation for SAS's adaptive sampling.
}

\textbf{Fitting.}
PROBE fits $a$, $c$ and $c_L$ together by weighted maximum likelihood, as in Platt scaling~\cite{platt1999probabilistic}. PROBE minimizes it over both samples,

\begin{equation}
\label{eq:loss}
\min_{a\ge 0,\,c,\,c_L}\;
 \sum_{i\in A}
 \frac{\omega_i}{\bar\omega}\,
 \mathrm{BCE}(y_i^{\mathrm{LLM}},g_i)
 +
 \sum_{i\in L}\mathrm{BCE}(y_i^{\mathrm{LLM}},g_i),
\end{equation}
where
$\mathrm{BCE}(y,g)=-y\log g-(1-y)\log(1-g)$,
$\omega_i$ is the inverse of the probability that row $i$
was drawn into $A$, i.e., the number of rows of $U$ it stands for,
and $\bar\omega$ is the mean weight over $A$, so that $A$ counts as $|A|$ rows.
Weighting the log-likelihood of each row by the inverse of its inclusion probability is the standard way to fit a model on an unequal-probability sample~\cite{manski1977estimation,binder1983variances}.
For the rows of $L$, PROBE uses out-of-fold features:
it splits $L$ into $R_{\mathrm{fold}}$ stratified folds and,
for each fold, retrains the proxy and reruns graph refinement
without it, yielding $f_i^{\mathrm{oof}}$ on the held-out rows. 
This prevents calibration on training scores that were computed using the candidates' own LLM labels.

\textbf{{Weighted fit on calibration sample set $A$.}}
{}
For a fixed shared slope $a$, the intercept $c$ is determined
by the inverse-probability-weighted loss on $A$.
For $m>0$, every row in $U_j$ has inclusion probability
$n_j/N_j>0$. Conditional on $L$, the Horvitz--Thompson
estimate~\cite{horvitz1952generalization}
$\widehat N^{+}=\sum_{i\in A}\omega_i y_i^{\mathrm{LLM}}$
is unbiased for the number of LLM-positive rows in $U$
under the calibration sampling.
At a finite optimum of Eq.~\eqref{eq:loss},
setting its derivative with respect to $c$ to zero gives
{}
\begin{equation}
\label{eq:anchor}
\sum_{i\in A}\omega_i\,
\operatorname{sig}\!\left(a\,{\operatorname{logit}_{\epsilon}}(f_i)+c\right)
=
\widehat{N}^{+}
=
\sum_{i\in A}\omega_i\,y_i^{\mathrm{LLM}}.
\end{equation}
{}
This equality balances the weighted predicted and observed
positive counts on the calibration sample $A$.
It is the weighted form of a known property of logistic regression: with an intercept, the fitted and the observed numbers of positives are equal.
PROBE routes with
$p_i=\operatorname{sig}(a\,{\operatorname{logit}_{\epsilon}}(f_i)+c)$
and drops $c_L$.
When $A$ contains no positive sample, $\widehat N^{+}=0$,
and Eq.~\eqref{eq:anchor} would set every $p_i$ to zero.
PROBE then uses an upper bound to set a target number of positives in $U$ and fits $c$ to this target.
It fits $a$ and $c_L$ on $L$ alone,
computes a one-sided Clopper--Pearson upper bound~\cite{clopper1934} at level $\delta_{\mathrm{CP}}=0.05$ on the number of positives in $U$,
and solves Eq.~\eqref{eq:anchor} for $c$
with $\widehat N^{+}$ replaced by a heuristic share $\nu_{\mathrm{CP}}=0.15$ of this bound.
Calibration fails if the solver for Eq.~\eqref{eq:loss} does not converge. PROBE then keeps the refined labels without calibration.

\subsection{Routing Objective}
\label{sec:calib:routing}

After calibration, the candidates in $S=L\cup A$ have LLM labels, while each remaining candidate $x_i\notin S$ has a calibrated match probability $p_i$ and a refined label $y_i^{\mathrm{refined}}$. 
PROBE selects an additional set $D\subseteq\mathcal X\setminus S$ for LLM evaluation.
As defined in Section~\ref{sec:bg:problem}, the final output uses LLM labels for candidates in $S\cup D$ and refined
labels for the others.
To estimate output quality before evaluating $D$, PROBE uses $p_i$ to estimate the true-positive, false-positive, and false-negative counts relative to exhaustive LLM evaluation:

\begin{equation}
\label{eq:counts}
\begin{split}
\widetilde{\mathit{TP}}(D)
&=
C+
\textstyle\sum_{i\notin S,\,y_i^{\mathrm{refined}}=1}p_i
+
\sum_{i\in D,\,y_i^{\mathrm{refined}}=0}p_i,
\\
\widetilde{\mathit{FP}}(D)
&=
\textstyle\sum_{i\notin S\cup D,\,
y_i^{\mathrm{refined}}=1}(1-p_i),
\\
\widetilde{\mathit{FN}}(D)
&=
\textstyle\sum_{i\notin S\cup D,\,
y_i^{\mathrm{refined}}=0}p_i,
\end{split}
\end{equation}
where $C=\sum_{x_i\in S}y_i^{\mathrm{LLM}}$ counts the known positives in $S$.
The estimated output quality is
$$\widetilde F_1(D)=
\frac{2\widetilde{\mathit{TP}}(D)}{
2\widetilde{\mathit{TP}}(D)+
\widetilde{\mathit{FP}}(D)+
\widetilde{\mathit{FN}}(D)}$$

Replacing the unknown $F_1$ in Eq.~\eqref{eq:problem} with this estimate, PROBE selects $D$ by minimizing
\begin{equation}
\label{eq:objective}
\min_{D\subseteq\mathcal X\setminus S}
\;\;
\alpha\underbrace{\bigl(1-\widetilde F_1(D)\bigr)}
_{\text{estimated quality loss}}
+ \;\;
(1-\alpha)\underbrace{d(D)}
_{\text{delegation rate}}.
\end{equation}

\noindent\textbf{Routing set.}
Verifying a predicted positive removes an expected $1-p_i$ false positives, while verifying a predicted negative converts an expected $p_i$ false negatives into true positives.
Since the calibrated probabilities $p_i$ are non-decreasing in refined scores $f_i$ ($a\geq0$), PROBE prioritizes predicted positives with the lowest scores and predicted negatives with the highest scores.

Let $i_1^+,\ldots,i_{m^+}^+$ index the predicted positives outside $S$ in increasing order of $f_i$,
and $i_1^-,\ldots,i_{m^-}^-$ index the predicted negatives in decreasing order.
An optimal routing set therefore has the form
\begin{equation}
\label{eq:two-threshold}
D^*=
\{x_{i_j^+}:1\le j\le r^+\}
\;\cup\;
\{x_{i_j^-}:1\le j\le r^-\},
\end{equation}
for some $r^+$ and $r^-$.

For a fixed number $r=r^++r^-$ of routed candidates, the delegation cost is constant depending only on $r=r^++r^-$, so PROBE picks the split $r^+$ that maximizes $\widetilde F_1$.
The numerator and denominator of $\widetilde F_1$ are linear in the $p_i$ of the selected candidates, so PROBE uses Dinkelbach's method~\cite{dinkelbach1967} to transform this ratio optimization into a sequence of objectives concave in $r^+$, each solved by binary search.
It then selects the best $r$ to minimize Eq.~\eqref{eq:objective}.

\begin{figure*}[t]
\captionsetup{skip=2pt}
\centering
\includegraphics[width=\textwidth]{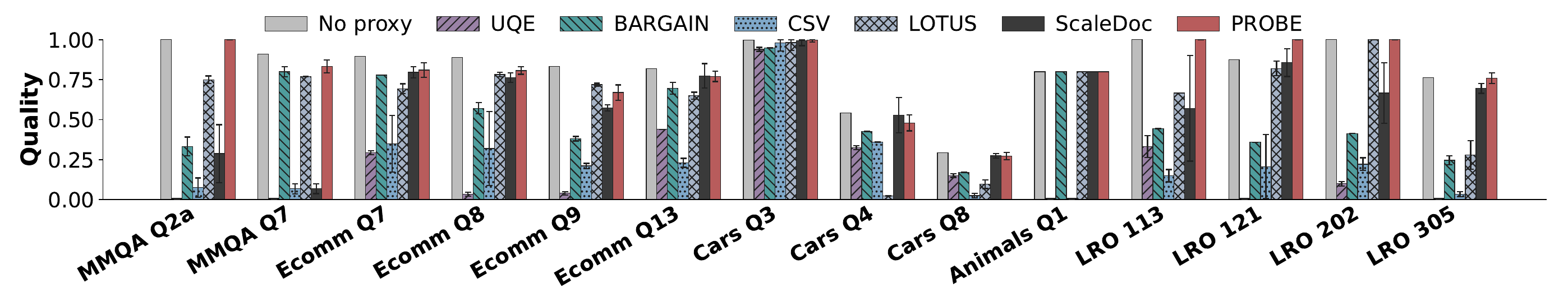}
\caption{Mean $F_1$ over five runs on the evaluated queries. Error bars indicate ±1 standard deviation across seeds. Budget-based systems use a $10\%$ sampling budget and target-based systems use a target of $0.9$.}
\label{fig:sys-f1}
\vspace{2pt}
\includegraphics[width=\textwidth]{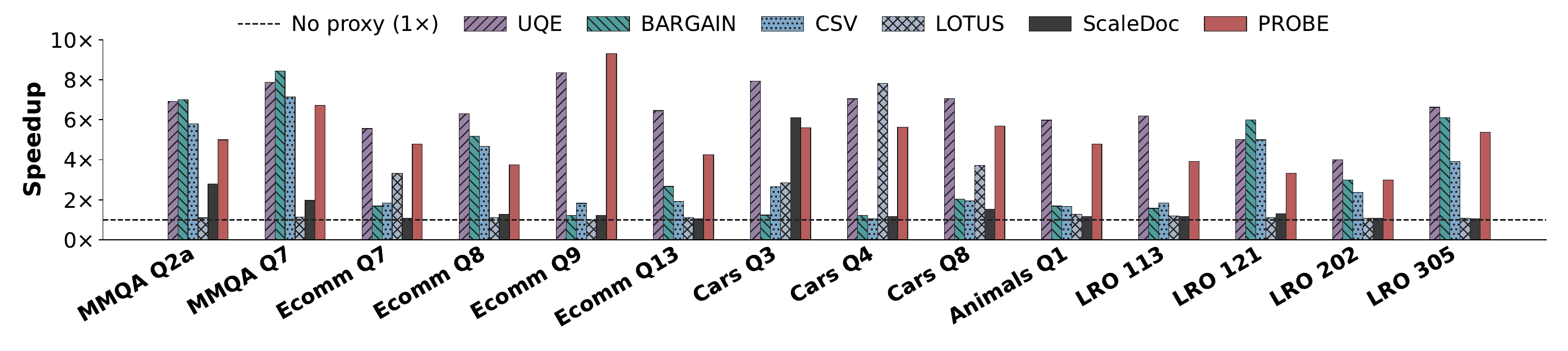}
\vspace{-5mm}
\caption{Speedup over the no-proxy baseline under the settings of \autoref{fig:sys-f1}.}
\label{fig:sys-latency}
\vspace{-1mm}
\end{figure*}

\section{Evaluation}
\label{sec:eval}

\subsection{Experimental Setup}
\label{sec:eval:setup}

\textbf{Implementation and configuration.}
PROBE is implemented on a SparkSQL/vLLM stack. Unless otherwise stated, all systems use Gemma~4 26B-A4B~\cite{gemma4} as the expert LLM. All systems send their LLM calls to the same vLLM server with the same batching configuration. All experiments run on one server with two NVIDIA RTX PRO 6000 Blackwell Max-Q GPUs, a 20-core Intel Xeon w7-3445 and 128\,GB of DRAM. SAS uses the default parameters of \secref{sec:sampling:algo}, and \autoref{tab:hparams-main} lists those of the later stages.

\begin{table}[t]
\centering
\footnotesize
\setlength{\tabcolsep}{3pt}
\caption{Default parameters of proxy training, graph refinement, calibration and routing.}
\label{tab:hparams-main}
{
\begin{tabularx}{\columnwidth}{@{}lXl@{}}
\toprule
Symbol & Meaning & Default \\
\midrule
$C_{\mathrm{LR}}$ & inverse $L_2$ strength of the logistic-regression proxy & $1$ \\
$k$ & neighbors per row in the $k$-NN graph & \addr{$\lfloor 3\log_2 N\rceil$} \\
$\beta,\gamma$ & weight of bias diffusion and of homogeneity propagation & $0.8,\,0.8$ \\
$T_1,T_2$ & maximum iterations of the two refinement phases & $50$ \\
$\theta_{\mathrm{ref}}$ & threshold that turns refined scores into labels & $0.5$ \\
$\kappa_L$ & share of the sampling budget for proxy-training labels{, }the rest is for calibration & $0.6$ \\
$n_{\mathrm{bin}}$ & equal-width refined-score strata for calibration & $10$ \\
$\varepsilon_A$ & share of calibration draws split evenly across strata (at least one per stratum) & $0.2$ \\
$R_{\mathrm{fold}}$ & cross-fitting folds{, }$n_\pm$: positive/negative training labels & $\min(5,n_+,n_-)$ \\
{$\epsilon$} & {score clipping level before the calibration logit} & {$10^{-6}$} \\
{$\nu_{\mathrm{CP}},\delta_{\mathrm{CP}}$} & {share and level of the Clopper--Pearson target (\secref{sec:calib:model})} & $0.15,\,0.05$ \\
$\alpha$ & quality--cost tradeoff of routing & $0.5$ \\
\bottomrule
\end{tabularx}}
\end{table}

\noindent\textbf{Datasets and queries.} We evaluate PROBE on the queries with the lowest ground-truth positive rates across SemBench~\cite{lao2026sembench} and LRO-Bench~\cite{lro}. The workload mixes semantic filters and joins over text and image inputs, with positive rates ranging from $2.17{\times}10^{-5}$ to $1.39{\times}10^{-1}$ and rows from a few hundred to over half a million as shown in \autoref{tab:workloads}. PROBE's routing objective measures $F_1$ against the expert LLM labels (\secref{sec:calib:routing}), since these are the only labels available during execution. The $F_1$ values reported in this section are instead measured against the human-annotated ground-truth labels.

\begin{figure*}[t]
\centering
\includegraphics[width=\textwidth]{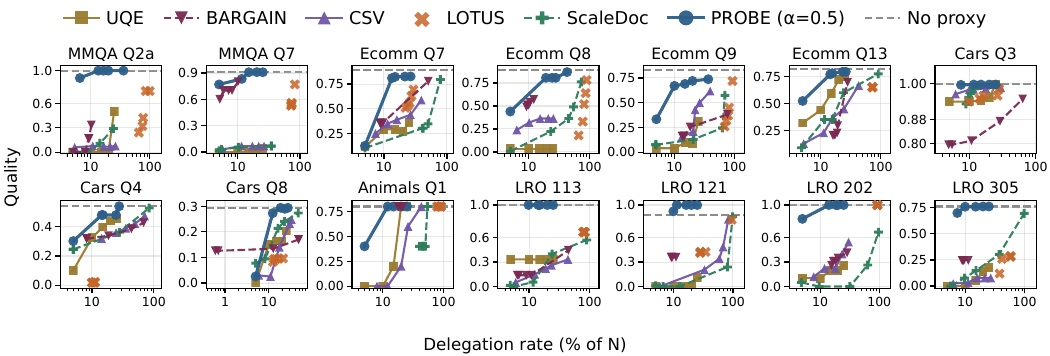}
\vspace{-7mm}
\caption{Quality--cost tradeoff of each system. The grey dashed line marks the quality of full LLM evaluation.}
\label{fig:sys-tradeoff}
\vspace{-1mm}
\end{figure*}

\begin{table}[!t]
\centering\small
\setlength{\tabcolsep}{3pt}
\caption{The 14 evaluated queries. Type F is a semantic filter, and J is a semantic join, and Size is the number of candidates (tuples for filters, tuple pairs for joins).}
\vspace{-3mm}
\label{tab:workloads}
{\footnotesize

\begin{tabular}{@{}llllrr@{}}
\toprule
Query & Scenario & Type & Modality & Size & Positive Rate \\
\midrule
MMQA Q2a & MMQA       & J & text+image      & $10{,}400$    & $5{\times}10^{-4}$ \\
MMQA Q7 & MMQA       & J & text+image      & $230{,}400$ & $2.17{\times}10^{-5}$ \\
Ecomm Q7 & E-commerce & J & text       & $525{,}625$   & $9.4{\times}10^{-3}$ \\
Ecomm Q8 & E-commerce & J & text+image      & $115{,}768$   & $3{\times}10^{-4}$ \\
Ecomm Q9 & E-commerce & J & text+image      & $25{,}440$    & $2.65{\times}10^{-2}$ \\
Ecomm Q13 & E-commerce & F & text+image & $3{,}992$     & $3{\times}10^{-3}$ \\
Cars Q3 & Cars       & F & text+image      & $10{,}070$    & $3.39{\times}10^{-2}$ \\
Cars Q4 & Cars       & F & text       & $157{,}269$   & $1.39{\times}10^{-1}$ \\
Cars Q8 & Cars       & F & text+image      & $30{,}130$    & $5.4{\times}10^{-3}$ \\
Animals Q1 & Animals    & F & text+image      & $1{,}600$      & $6.1{\times}10^{-2}$ \\
LRO 113 & Santos   & J & text       & $266$       & $7.52{\times}10^{-3}$ \\
LRO 121 & Santos   & J & text       & $1{,}426$     & $4.91{\times}10^{-3}$ \\
LRO 202 & Movie   & J & text       & $384$       & $7.81{\times}10^{-3}$ \\
LRO 305 & Schools  & J & text       & $1{,}421$     & $7.74{\times}10^{-3}$ \\
\bottomrule
\end{tabular}
}
\end{table}

\noindent\textbf{Baselines.} We compare PROBE against five recent proxy based semantic query systems and the no-proxy baseline\condr{, which is the PROBE implementation that disables all optimizations and invokes the LLM on every input row}. All systems use the same prompts. PROBE uses the scores of the Qwen3-VL-Reranker-2B~\cite{qwen3vlreranker} cross-encoder to rank rows for sampling.

\begin{itemize}[leftmargin=*, itemsep=2pt, topsep=2pt]

\item \textbf{LOTUS}~\cite{patel2025lotus} optimizes semantic operators with model cascades and accuracy-aware approximation. For filters, its proxy is the token log-probability of Qwen3-VL-2B~\cite{qwen3vl}. For joins, we enable LOTUS's join cascade, whose proxy is the embedding similarity of e5-base-v2 for text and CLIP ViT-B/32 for images. We set both its recall and precision targets to $0.9$, with failure probability $\delta=0.1$ and a default sampling rate of $10\%$.

\item \textbf{BARGAIN}~\cite{bargain} cascades a cheap proxy before the LLM and calibrates a confidence threshold to meet a SUPG-style~\cite{supg} statistical guarantee. We use its accuracy mode with target $0.9$ and failure probability $\delta=0.1$, with no cap on LLM budget. \condr{It uses the cosine similarity between the row and predicate embeddings as the proxy score, computed with the same embedding models as LOTUS's join proxy.}

\item \textbf{ScaleDoc}~\cite{zhang2026scaledoc} trains a query-specific proxy, an MLP projection over the frozen embeddings trained with contrastive learning, and calibrates a two-threshold cascade that sends rows between the thresholds to the LLM. We use $5\%$/$5\%$ of the rows for training/calibration and an $F_1$ target of $0.9$.

\item \textbf{CSV}~\cite{hou2026csv} clusters the embeddings with $k$-means ($k=4$, and $k=2$ when re-clustering), samples $10\%$ of each cluster for LLM labels, and assigns a cluster's label when at least $90\%$ of its sampled rows agree.

\item \textbf{UQE}~\cite{uqe} is the sampling and proxy training method that AIQA~\cite{chung-proxy} uses. It spends the budget $B$ with online active learning: in each round, it labels the unlabeled rows that the current logistic-regression proxy scores highest and refits the proxy on all labeled rows. The final proxy labels all remaining rows.

\end{itemize}

\subsection{End-to-end Comparison}
\label{sec:eval:main}

\autoref{fig:sys-f1} and \autoref{fig:sys-latency} compare the systems under the defaults of \secref{sec:eval:setup}: the budget-based systems (UQE, CSV and PROBE) use $B/N=10\%$, with $\alpha=0.5$ for PROBE, and the target-based systems (LOTUS, BARGAIN and ScaleDoc) use a target of 0.9. \condr{The two limitations of \autoref{sec:intro} both appear. LOTUS and ScaleDoc over-delegate yet achieve lower quality: ScaleDoc reaches a geometric-mean $F_1$ of 0.530 at a geometric-mean delegation rate of 59.8\%, and LOTUS sends 54.6\% of the rows to the LLM for an $F_1$ of 0.485. BARGAIN delegates 25.6\%, but its accuracy target is met mostly by the many non-matching rows, so its geometric-mean $F_1$ stays at 0.473. The budget-based systems delegate less but lose quality: UQE and CSV reach $F_1$ below 0.2 on 9 and 6 queries, respectively.}
PROBE attains a geometric-mean $F_1$ of 0.762, compared to 0.797 for the no-proxy baseline, with a geometric-mean delegation rate of 12.8\%, a 4.87$\times$ geometric-mean speedup and a maximum speedup of 9.3$\times$.

Because several queries have very few positives, we rerun each system's configuration with five seeds and test whether PROBE's $F_1$ differs from that of each baseline. We draw bootstrap samples by resampling the seeds and, within each query, the positive and negative rows separately, and compute the 95\% confidence interval (CI) for each mean $F_1$ and each difference. A difference is significant at the $0.05$ level when its 95\% CI excludes zero. Over the 14 queries, PROBE's geometric mean $F_1$ is $0.762$ (95\% CI [$0.733$, $0.789$]), and it significantly outperforms all five proxy-based baselines. On the seven queries with fewer than 20 positives, PROBE reaches an $F_1$ of $0.902$ (95\% CI [$0.848$, $0.920$]), compared to $0.905$ [$0.865$, $0.961$] for the full LLM evaluation.
These results show that \textbf{PROBE can approach exhaustive LLM quality under extreme positive scarcity without evaluating most candidates with the LLM.}

\subsection{Quality--Cost Tradeoff}
\label{sec:eval:tradeoff}
\begin{table*}[t]
\centering
\footnotesize
\setlength{\tabcolsep}{2pt}
\renewcommand{\arraystretch}{1.10}
\definecolor{ProbeBestGreen}{RGB}{153,211,149}
\definecolor{ProbeSecondGreen}{RGB}{193,229,187}
\definecolor{ProbeFastBlue}{RGB}{205,228,247}
\newcommand{\probeFBest}[1]{\cellcolor{ProbeBestGreen}\textbf{#1}}
\newcommand{\probeFSecond}[1]{\cellcolor{ProbeSecondGreen}#1}
\newcommand{\probeFast}[1]{\cellcolor{ProbeFastBlue}\textbf{#1}}
\caption{Best operating points from \secref{sec:eval:tradeoff}, averaged over five seeds. Each configuration uses the lowest-$d$ point attaining its highest $F_1$; PROBE uses $\alpha=0.9$ (quality-first) or $\alpha=0.1$ (cost-first). Cells with dark green bold and light green indicate the best and second-best $F_1$, and blue bold indicates the lowest latency; ties are broken by the other metrics. Delegation $d$ is in percent and latency is in seconds ($k$ denotes thousands). The bottom rows report geometric means and standard deviations across seeds.}
\label{tab:main-comparison}
\vspace{-3mm}
\resizebox{\textwidth}{!}{%
\begin{tabular}{llrrrrrrrrrrrrrrrrrrrrrrr}
\toprule
\multirow{2}{*}{Dataset} & \multirow{2}{*}{Query} & \multicolumn{2}{c}{No proxy} & \multicolumn{3}{c}{LOTUS} & \multicolumn{3}{c}{BARGAIN} & \multicolumn{3}{c}{CSV} & \multicolumn{3}{c}{ScaleDoc} & \multicolumn{3}{c}{UQE} & \multicolumn{3}{c}{\begin{tabular}{@{}c@{}}PROBE\\Quality-first\end{tabular}} & \multicolumn{3}{c}{\begin{tabular}{@{}c@{}}PROBE\\Cost-first\end{tabular}} \\
\cmidrule(lr){3-4}\cmidrule(lr){5-7}\cmidrule(lr){8-10}\cmidrule(lr){11-13}\cmidrule(lr){14-16}\cmidrule(lr){17-19}\cmidrule(lr){20-22}\cmidrule(lr){23-25}
 & & $F_1\!\uparrow$ & Lat.$\downarrow$ & $F_1\!\uparrow$ & $d\!\downarrow$ & Lat.$\downarrow$ & $F_1\!\uparrow$ & $d\!\downarrow$ & Lat.$\downarrow$ & $F_1\!\uparrow$ & $d\!\downarrow$ & Lat.$\downarrow$ & $F_1\!\uparrow$ & $d\!\downarrow$ & Lat.$\downarrow$ & $F_1\!\uparrow$ & $d\!\downarrow$ & Lat.$\downarrow$ & $F_1\!\uparrow$ & $d\!\downarrow$ & Lat.$\downarrow$ & $F_1\!\uparrow$ & $d\!\downarrow$ & Lat.$\downarrow$ \\
\midrule
\multirow{2}{*}{MMQA} & Q2a & 1.000 & 1.13k & 0.750 & 85.5 & 1.03k & 0.333 & 10.3 & 161.2 & 0.075 & 10.9 & 194.6 & 0.288 & 24.4 & 404.2 & 0.500 & 25.0 & 433.0 & \probeFBest{1.000} & 13.5 & 225.9 & \probeFSecond{0.909} & 6.6 & \probeFast{124.9} \\
 & Q7 & 0.909 & 4.30k & 0.769 & 82.9 & 3.76k & 0.800 & 10.1 & \probeFast{509.3} & 0.068 & 11.5 & 601.1 & 0.068 & 34.7 & 2.19k & 0.040 & 25.0 & 1.35k & \probeFBest{0.909} & 15.1 & 810.6 & \probeFSecond{0.833} & 11.6 & 640.4 \\
\midrule
\multirow{4}{*}{Ecomm} & Q7 & 0.895 & 6.79k & 0.694 & 29.4 & 2.04k & 0.778 & 50.5 & 4.03k & 0.588 & 39.2 & 3.10k & 0.796 & 78.9 & 6.30k & 0.364 & 25.0 & 2.07k & \probeFBest{0.827} & 21.5 & 2.29k & \probeFSecond{0.810} & 13.4 & \probeFast{1.42k} \\
 & Q8 & 0.889 & 5.24k & 0.782 & 89.0 & 4.68k & 0.571 & 12.1 & 1.01k & 0.364 & 15.4 & 1.35k & 0.762 & 73.0 & 4.10k & 0.034 & 5.0 & \probeFast{573.2} & \probeFBest{0.869} & 42.6 & 3.03k & \probeFSecond{0.807} & 19.2 & 1.40k \\
 & Q9 & 0.833 & 438.1 & 0.719 & 93.0 & 436.2 & 0.381 & 76.5 & 358.7 & 0.617 & 39.5 & 174.1 & 0.574 & 69.5 & 358.0 & 0.308 & 25.0 & \probeFast{143.3} & \probeFSecond{0.737} & 36.1 & 178.3 & \probeFBest{0.737} & 36.0 & 177.4 \\
 & Q13 & 0.818 & 316.9 & 0.650 & 73.0 & 284.8 & 0.696 & 28.0 & 118.7 & 0.667 & 42.8 & 165.5 & \probeFSecond{0.774} & 91.2 & 304.4 & 0.750 & 25.0 & 110.3 & \probeFBest{0.793} & 23.2 & 93.9 & 0.769 & 14.4 & \probeFast{74.5} \\
\midrule
\multirow{3}{*}{Cars} & Q3 & 0.998 & 711.9 & 0.982 & 28.2 & 249.9 & 0.949 & 63.8 & 574.6 & 0.987 & 29.7 & 267.1 & 0.993 & 13.5 & 116.4 & 0.996 & 25.0 & 216.1 & \probeFBest{0.997} & 11.7 & 127.3 & \probeFSecond{0.996} & 7.5 & \probeFast{87.8} \\
 & Q4 & 0.542 & 1.99k & 0.023 & 11.0 & \probeFast{254.9} & 0.427 & 68.4 & 1.63k & 0.473 & 71.0 & 1.88k & 0.528 & 83.9 & 1.70k & 0.452 & 25.0 & 582.0 & \probeFSecond{0.540} & 27.5 & 586.2 & \probeFBest{0.540} & 26.5 & 569.4 \\
 & Q8 & 0.294 & 1.83k & 0.096 & 17.0 & 490.0 & 0.170 & 47.7 & 895.3 & 0.248 & 32.8 & 932.7 & \probeFSecond{0.274} & 47.0 & 1.18k & 0.203 & 25.0 & 708.6 & \probeFBest{0.294} & 18.6 & 473.2 & 0.273 & 12.3 & \probeFast{321.2} \\
\midrule
Animals & Q1 & 0.800 & 77.6 & 0.800 & 78.0 & 60.8 & 0.800 & 18.0 & 46.0 & 0.800 & 43.0 & 52.8 & 0.800 & 54.0 & 65.9 & 0.800 & 20.0 & 24.1 & \probeFSecond{0.800} & 12.0 & 16.2 & \probeFBest{0.800} & 11.0 & \probeFast{16.0} \\
\midrule
\multirow{4}{*}{LRO} & 113 & 1.000 & 4.4 & 0.667 & 78.0 & 3.7 & 0.444 & 44.7 & 2.8 & 0.333 & 43.9 & 2.4 & 0.571 & 89.3 & 3.8 & 0.333 & 5.0 & \probeFast{0.3} & \probeFBest{1.000} & 9.8 & 0.8 & \probeFSecond{1.000} & 9.8 & 0.8 \\
 & 121 & 0.875 & 3.0 & 0.824 & 96.0 & 2.7 & 0.359 & 9.1 & \probeFast{0.5} & 0.824 & 83.8 & 2.7 & 0.857 & 99.1 & 2.3 & 0.111 & 25.0 & 1.1 & \probeFBest{1.000} & 11.3 & 0.9 & \probeFSecond{0.923} & 9.9 & 0.7 \\
 & 202 & 1.000 & 1.2 & 1.000 & 93.0 & 1.1 & 0.414 & 28.5 & 0.4 & 0.545 & 30.3 & 0.6 & 0.667 & 97.0 & 1.1 & 0.261 & 25.0 & 0.4 & \probeFBest{1.000} & 14.0 & \probeFast{0.4} & \probeFSecond{1.000} & 14.0 & 0.4 \\
 & 305 & 0.762 & 5.5 & 0.286 & 58.0 & 5.1 & 0.246 & 8.9 & 0.9 & 0.078 & 15.9 & 1.4 & 0.696 & 98.0 & 5.2 & 0.176 & 25.0 & 1.9 & \probeFBest{0.762} & 19.9 & 2.0 & \probeFSecond{0.700} & 7.4 & \probeFast{0.8} \\
\midrule
\multicolumn{2}{l}{\textbf{Geo. mean}} & 0.797 & 213.9 & 0.485 & 54.6 & 132.0 & 0.473 & 25.6 & 77.4 & 0.354 & 30.7 & 88.8 & 0.530 & 59.8 & 145.6 & 0.262 & 19.6 & 58.8 & \probeFBest{0.790} & 17.9 & 59.5 & \probeFSecond{0.760} & 12.7 & \probeFast{43.8} \\
\multicolumn{2}{l}{\textbf{Std. dev.}} & -- & -- & 0.088 & 6.2 & 11.1 & 0.010 & 0.2 & 2.5 & 0.087 & 1.8 & 6.9 & 0.066 & 3.9 & 10.2 & 0.042 & 0.0 & 4.7 & 0.051 & 0.8 & 4.7 & 0.012 & 0.2 & 3.3 \\
\bottomrule
\end{tabular}%
}
\end{table*}

We sweep target values over $\{0.1,0.3,0.5,0.7,0.9\}$ for the target-based systems, and sampling budgets over $B/N\in\{5,10,15,20,25\}\%$ for the budget-based systems. \autoref{fig:sys-tradeoff} shows the quality-cost tradeoff curves of each system, and \autoref{tab:main-comparison} reports selected best operating points. For each baseline it reports the lowest-$d$ point among those that reach the baseline's highest quality on the query. For PROBE, the \emph{quality-first} and \emph{cost-first} points are selected from separate budget sweeps with $\alpha=0.9$ and $\alpha=0.1$, respectively. At each fixed $\alpha$, we sweep the sampling budget and, for each query, select the lowest-$d$ point attaining the highest $F_1$ in the budget sweep. These two operating-point configurations are separate from the default setting of $\alpha=0.5$ and $B/N=10\%$.

\autoref{fig:sys-tradeoff} shows that across the delegation range, PROBE's curve lies above or on the baselines' on most queries. As \autoref{tab:main-comparison} shows, at its quality-first points, PROBE attains a geometric-mean quality of $0.790$, against $0.797$ for the no-proxy baseline, while sending $17.9\%$ of the rows to the LLM. Its geometric-mean latency is $59.5$\,s against $213.9$\,s, a $3.6\times$ speedup. The strongest baseline, ScaleDoc, only reaches $0.530$ while sending $59.8\%$ of the rows to the LLM. At its cost-first points, PROBE keeps at least $90\%$ of the no-proxy quality except on Ecomm Q9 and achieves a geometric-mean quality of $0.760$ at $12.7\%$ delegation. Its geometric-mean latency is $43.8$\,s, corresponding to a $4.9\times$ speedup.

\subsection{Ablation on Sampling Policy}
\label{sec:eval:ablation-sampling}

\begin{figure}[t]
\centering
\begin{minipage}[t]{0.5\columnwidth}
\centering
\includegraphics[width=\linewidth,keepaspectratio]{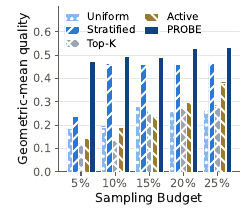}\\[-1pt]
{\footnotesize (a) Sampling policy}
\end{minipage}\hfill
\begin{minipage}[t]{0.5\columnwidth}
\centering
\includegraphics[width=\linewidth,keepaspectratio]{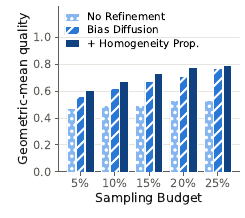}\\[-1pt]
{\footnotesize (b) Graph refinement stage}
\end{minipage}
\vspace{-3mm}
\caption{Geometric-mean quality for (a) five sampling policies with graph refinement disabled and (b) the graph refinement stages added on top of PROBE sampling.}
\label{fig:ablation}
\end{figure}

We isolate the sampling policy by disabling graph refinement, calibration, and routing. Each method uses the same training-label budget and trains the same logistic-regression proxy on its acquired LLM labels, and then applies the proxy to predict the remaining candidates. 
We compare PROBE’s SAS with uniform sampling, stratified sampling (Neyman allocation), top-$K$ sampling, and UQE's active sampling~\cite{uqe}, varying the training-label budget $|L|/N$ from $5\%$ to $25\%$.
As \autoref{fig:ablation}(a) shows, PROBE achieves the highest geometric-mean $F_1$ at every budget.
Top-$K$ and active sampling perform worst at $5\%$ and $10\%$, while active sampling outperforms uniform and top-$K$ sampling only at $20\%$ and $25\%$. The results highlight the difficulty and importance of acquiring sufficient positive samples under high selectivity to train an accurate proxy, independently of subsequent proxy refinement and routing.

\subsection{Ablation on Graph Refinement Stage}
\label{sec:eval:ablation-refinement}

We fix the labeled set acquired by PROBE’s SAS and incrementally enable the two graph refinement stages, bias diffusion and homogeneity propagation (\secref{sec:ssl}), using the unrefined logistic-regression proxy as the baseline. 
As \autoref{fig:ablation}(b) shows, graph refinement improves geometric-mean $F_1$ from $0.47$--$0.53$ to $0.60$--$0.79$ across sampling budgets, with a gain of $1.3$--$1.5\times$.
Bias diffusion accounts for most of the improvement, while homogeneity propagation adds at most $0.07$.
The larger contribution of bias diffusion is consistent with the local error correlation assumption in \secref{sec:ssl:objective}.

\begin{figure}[t]
\centering
\begin{minipage}[t]{0.48\columnwidth}
\centering\vspace{0pt}
\includegraphics[width=\linewidth,keepaspectratio]{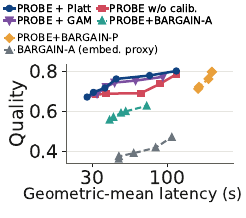}
\caption{Quality ($F_1$) versus latency as $\alpha$ or the BARGAIN target (0.1--0.9) varies.}
\label{fig:alpha-tradeoff}
\end{minipage}\hfill
\begin{minipage}[t]{0.48\columnwidth}
\centering\vspace{0pt}
\includegraphics[width=\linewidth,keepaspectratio]{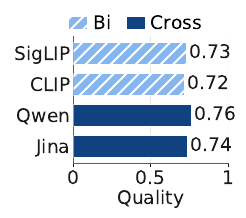}
\caption{Geometric-mean quality across different embedding models.}
\label{fig:repr-models}
\end{minipage}
\end{figure}

\subsection{Ablation on Learned Calibrated Cascade Routing}
\label{sec:eval:ablation-calibration}

\autoref{fig:alpha-tradeoff} compares three PROBE variants with the same sampling, graph refinement, and routing algorithm, but different calibration methods: Platt calibration (the default), the GAM calibrator~\cite{liskowski2026streaming}, and no calibration, where routing uses uncalibrated refined scores.
At the default $\alpha=0.5$, Platt calibration improves geometric-mean $F_1$ from $0.691$ to $0.762$ at a geometric-mean latency of $43.9$\,s, while reducing expected calibration error (ECE) from $0.213$ to $0.057$. 
GAM reaches an $F_1$ of $ 0.744$ at the same $\alpha$ and remains below Platt calibration at all higher tested values of $\alpha$. 
These improvements in both calibration error and output $F_1$ highlight the importance of reliable match probabilities for cost-aware routing.
Without routing ($\alpha=0$), PROBE achieves $0.672$ at $27.5$\,s. Further increasing $\alpha$ yields diminishing returns: compared with $\alpha=0.5$, $\alpha=0.9$ improves $F_1$ by about $5\%$ but increases latency by $2.6\times$.

To compare PROBE's calibrated routing with BARGAIN, we replace PROBE's calibration and routing with
BARGAIN~\cite{bargain}, while keeping the same refined scores. 
We denote its accuracy and precision modes by PROBE+BARGAIN-A and PROBE+BARGAIN-P, respectively.
PROBE+BARGAIN-A achieves a geometric-mean $F_1$ of $0.56$--$0.63$, outperforming BARGAIN-A with an embedding proxy in both quality and latency across all tested targets.
However, accuracy-based routing remains ineffective under high selectivity: the dominant negative class
allows BARGAIN's accuracy mode to meet its target despite missing positives.
Its precision mode does not obtain a certified threshold in these runs and relies heavily on LLM evaluation, limiting cost savings.
In contrast, PROBE directly optimizes estimated $F_1$ against LLM cost and achieves a better empirical quality--latency tradeoff than these BARGAIN variants.

\subsection{Sensitivity to Model Choice}
\label{sec:eval:model-sensitivity}
\label{sec:eval:embedding}
\paragraph{Representation model.}
PROBE uses one pretrained model to rank candidates for sampling (\secref{sec:sampling}) and to embed them for proxy training and graph construction~(\secref{sec:ssl}). We compare two bi-encoders, SigLIP~\cite{siglip} and CLIP-base~\cite{clip}, which embed the predicate and each candidate independently, with two multimodal cross-encoders, Qwen3-VL-Reranker-2B~\cite{qwen3vlreranker} and jina-reranker-m0~\cite{jinam0}, which jointly encode the predicate and the candidate. As \autoref{fig:repr-models} shows, Qwen3-VL-Reranker-2B attains the highest geometric-mean quality, $0.762$, against $0.74$ for jina-reranker-m0, $0.73$ for SigLIP and $0.72$ for CLIP-base. PROBE's quality varies by at most 0.04 across representation models.

\begin{figure}[t]
\centering
\begin{minipage}[t]{0.5\columnwidth}
\centering\vspace{0pt}
\includegraphics[width=\linewidth,keepaspectratio]{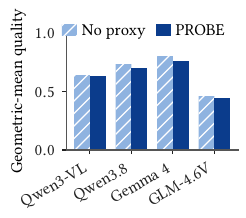}\\[-1pt]
{\footnotesize (a) Quality}
\end{minipage}\hfill
\begin{minipage}[t]{0.5\columnwidth}
\centering\vspace{0pt}
\includegraphics[width=\linewidth,keepaspectratio]{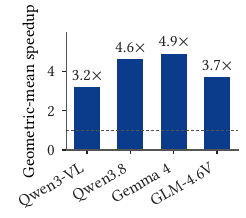}\\[-1pt]
{\footnotesize (b) Speedup}
\end{minipage}
\vspace{-3mm}
\caption{(a) Geometric-mean quality of the same-model no-proxy baseline and of PROBE. (b) Geometric-mean speedup of PROBE over the same-model no-proxy baseline.}
\label{fig:mllm-generalization}
\end{figure}
\label{sec:eval:mllm-model}

\paragraph{Expert LLM}
\autoref{fig:mllm-generalization} replaces the expert LLM with four open multimodal models: Qwen3-VL-30B-A3B-Instruct~\cite{qwen3vl}, Qwen3.8-27B~\cite{qwen38}, Gemma~4 26B-A4B~\cite{gemma4}, and GLM-4.6V-Flash~\cite{glm46v}. For each model, PROBE uses $B{/N}=10\%$ and $\alpha=0.5$ and is compared with the no-proxy baseline on the same model. Across the four models, PROBE's geometric-mean quality stays within $0.04$ of the no-proxy baseline, and it attains a geometric-mean speedup of $3.2$--$4.9\times$.

\begin{figure}[t]
\centering
\includegraphics[width=\columnwidth]{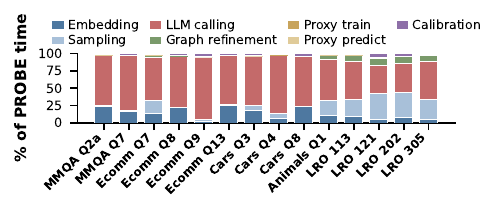}
\vspace{-5mm}
\caption{{Normalized end-to-end runtime breakdown of PROBE by query. \emph{LLM calling} covers the expert calls during sampling and cascade routing.}}
\label{fig:time-stack}
\end{figure}

\subsection{Time Breakdown}
\label{sec:eval:timing}

As \autoref{fig:time-stack} shows, LLM calling takes $72.1\%$ of PROBE's end-to-end time and embedding all rows $16.7\%$. Sampling ($6.6\%$), calibration ($2.3\%$), graph refinement with $k$-NN construction ($1.7\%$), and proxy training and prediction ($0.6\%$) add modest overhead. LLM calls remain the dominant cost, and embedding cost grows with the number of candidate pairs for large joins, and optimization such as blocking that prunes pairs before embedding is left to future work.

\subsection{Case Study: Less Selective Queries}
\label{sec:eval:lowsel}

\begin{table}[t]
\centering
\footnotesize
\setlength{\tabcolsep}{3pt}
\caption{$F_1$ and delegation rate $d$ on less selective queries ($\pi$: positive rate). Baselines use target $0.9$, PROBE $\alpha=0.5$.}
\label{tab:lowsel-results}
\begin{tabular}{@{}lrrrrrr@{}}
\toprule
& \multicolumn{2}{c}{{Cars Q7}} & \multicolumn{2}{c}{Ecomm Q6} & \multicolumn{2}{c}{Movie Q5} \\
& \multicolumn{2}{c}{\scriptsize {$N{=}30{,}130$, $\pi{=}0.147$}} & \multicolumn{2}{c}{\scriptsize $N{=}9{,}280$, $\pi{=}0.200$} & \multicolumn{2}{c}{\scriptsize $N{=}65{,}024$, $\pi{=}0.497$} \\
\cmidrule(lr){2-3}\cmidrule(lr){4-5}\cmidrule(l){6-7}
System & $F_1$ & $d$ & $F_1$ & $d$ & $F_1$ & $d$ \\
\midrule
No proxy & {$0.669$} & $100\%$  & $0.620$ & $100\%$  & $0.768$ & $100\%$ \\
BARGAIN-A  & {$0.643$} & {$40.2\%$} & $0.516$ & $13.5\%$ & $0.723$ & $78.0\%$ \\
ScaleDoc & {$0.612$} & {$41.7\%$} & $0.463$ & $11.7\%$ & $0.746$ & $86.9\%$ \\
\midrule
PROBE, $B{/N}=10\%$ & {$0.627$} & {$35.2\%$} & $0.435$ & $11.1\%$ & $0.650$ & $13.6\%$ \\
PROBE, $B{/N}=25\%$ & {$0.667$} & {$44.2\%$} & $0.620$ & $30.2\%$ & $0.710$ & $29.4\%$ \\
\bottomrule
\end{tabular}
\end{table}

\autoref{tab:lowsel-results} evaluates three less selective queries outside the main evaluation, with positive rates from
0.147 to 0.497. Increasing the sampling budget from $10\%$ to $25\%$
raises PROBE's $F_1$ from 0.627 to 0.667 on Cars Q7, from 0.435 to
0.620 on Ecomm Q6, and from 0.650 to 0.710 on Movie Q5. At the
larger budget, delegation rates range from $29.4\%$ to $44.2\%$.
For the less selective queries in our case study, a larger sampling budget is needed to bring PROBE closer to the quality of full LLM evaluation. This is due to the weaker separation between positive and negative rows in the representation space, which would limit the effectiveness of graph-based refinement and increase the need for LLM supervision.

\section{Related Work}
\label{sec:related}

\paragraph{AI query systems.} Prior semantic query systems such as LOTUS~\cite{patel2025lotus}, Palimpzest~\cite{liu2025palimpzest}, Abacus~\cite{russo2025abacus}, DocETL~\cite{docetl} and Stretto~\cite{sanmartino2026stretto} expose LLM operators and optimize them through model selection, cascades, indexes, or plan-level search. PROBE instead spends the LLM budget within a single operator.

\paragraph{Approximate query processing for AI query systems.}
SUPG~\cite{supg} samples proxy scores to return an approximate selection meeting a user-specified precision or recall target. ThalamusDB~\cite{thalamusdb} calibrates similarity thresholds from a small sample set with deterministic error bounds. Other systems use cluster-level voting or decomposition-based pruning~\cite{hou2026csv,fdj} and stratified allocation to estimate semantic-join aggregates~\cite{bas}. All of them expose a user-defined target, whereas GAMCAL~\cite{liskowski2026streaming} exposes a single cost--quality parameter. 

\paragraph{Learned routing and deferral.}
Model cascades and learned deferral decide whether an input is handled by a cheap proxy or escalated to an expert model
\cite{frugalgpt,hybridllm,routellm,gatekeeper,posthoc-defer,
jitkrittum2023deferral}. Prior LLM-routing work commonly treats each model request as the routing unit~\cite{huang2025thriftllm}, whereas a database query may apply a semantic operator to many rows. AIQA~\cite{chung-proxy} makes the routing decision for the entire database query. PROBE decides per row and routes only likely-wrong rows.
\section{Conclusion}
\label{sec:conclusion}

PROBE reduces LLM evaluations for highly selective semantic
queries by combining selectivity-adaptive sampling, graph-based
proxy refinement, and calibrated routing. On the evaluated queries, the default $10\%$ sampling budget yields a geometric-mean $F_1$ of $0.762$, compared with 0.797 for full LLM evaluation, while delegating 12.8\% of rows and achieving a $4.87\times$  geometric-mean speedup.

\bibliographystyle{ACM-Reference-Format}
\bibliography{sample}

\clearpage

\end{document}